\documentclass{article}

\usepackage[a4paper,top=3cm,bottom=2cm,left=3cm,right=3cm,marginparwidth=1.75cm]{geometry}

\usepackage[english]{babel}
\usepackage[utf8]{inputenc}
\usepackage[T1]{fontenc}
\usepackage{amsmath}
\usepackage{graphicx}
\usepackage{authblk}
\usepackage{xcolor}
\usepackage{soul}
\usepackage{float}
\usepackage{csquotes}
\usepackage[colorinlistoftodos]{todonotes}
\usepackage[colorlinks=true, allcolors=blue]{hyperref}
\usepackage{chngcntr}
\usepackage[modulo]{lineno}
\usepackage[backend=biber, style=nature]{biblatex} 
 \usepackage[modulo]{lineno}
\usepackage{enumitem} 
\usepackage{amssymb}
\usepackage{amsthm}
\usepackage{mathtools}
\theoremstyle{definition}
\newtheorem*{definition}{Definition}
\newtheorem*{remark}{Remark}
\newtheorem{theorem}{Theorem}
\newtheorem*{corollary}{Corollary}
\newtheorem{lemma}{Lemma}
\newtheorem{proposition}{Proposition}

\title{Uncovering complementary information sharing in spider monkey collective foraging using higher-order spatial networks}
\author[1,2]{Gabriel Ramos-Fernandez}
\author[3,4]{Ross S. Walker}
\author[4]{Matthew J. Silk}
\author[5]{Denis Boyer}
\author[1]{Sandra E. Smith-Aguilar}
\affil[1]{Instituto de Investigaciones en Matemáticas Aplicadas y en Sistemas, Universidad Nacional Autónoma de México, Mexico City, 04510 Mexico}
\affil[2]{Global Research Centre for Diverse Intelligences, University of St. Andrews, St Andrews, KY16 9AJ United Kingdom}
\affil[3]{Department of Mathematics and Maxwell Institute for Mathematical Sciences, Heriot-Watt University, Edinburgh, EH14 4AS United Kingdom}
\affil[4]{Institute of Ecology and Evolution, University of Edinburgh, Edinburgh, EH9 3FL United Kingdom} 
\affil[5]{Instituto de Física, Universidad Nacional Autónoma de México, Mexico City, 04510 Mexico}

\date{}

\begin{document}

\maketitle

\section*{Abstract}
Collectives are often able to process information in a distributed fashion, surpassing each individual member's processing capacity. In fission-fusion dynamics, where group members come together and split from others often, sharing complementary information about uniquely known foraging areas could allow a group to track a heterogenous foraging environment better than any group member on its own. We analyse the partial overlaps between individual spider monkey core ranges, which we assume represent the knowledge of an individual during a given season. Sets of individuals with complementary overlaps are identified, showing a balance between redundantly and uniquely known portions, and we use simplicial complexes to represent these higher-order interactions. The structure of the simplicial complexes shows holes in various dimensions, revealing complementarity in the foraging information that is being shared. We propose that the complex spatial networks arising from fission-fusion dynamics allow for adaptive, collective processing of foraging information in dynamic environments.

\newpage

\section*{Introduction}

A driver of the structure and dynamics of networked systems is their adaptive capacity to collectively process information in ways that none of the components would be able to alone \cite{mitchell2006complex}. A prevalent feature of this information processing is its distributed nature: different components complement one another in the information they contain, which they aggregate via the network of interactions. This complementarity in information aggregation is a hallmark of collectively intelligent systems  \cite{bettencourt2009rules, garg2025causal}.

In social animals, the sharing of information about resource availability between group members has long been considered one of the benefits of living in groups \cite{clark1984foraging, galef2001social, valone2002public}. Information transfer between individuals allows the group as a whole to gain better knowledge of the location of available food sources than any individual would gain on its own, especially in heterogenous, dynamic environments \cite{clark1984foraging, kummer2006primate}. This information transfer need not be through the use of active signals, and may simply require that naïve individuals follow knowledgeable ones to find food sources that the latter know uniquely (\textit{e.g.} \cite{sonerud2001ignorant, petit2010decision, king2011next}). In species where group members disperse over wide foraging areas, knowledge could be distributed among all of them, such that each may act as naïve or knowledgeable with respect to different food sources \cite{sonerud2001ignorant, palacios2019uncovering}. Sharing foraging information thus may allow the benefits of being in a group to outweigh the costs of competition caused by foraging with others \cite{clark1984foraging}.

Fission-fusion dynamics are a property of many animal groups \cite{aureli2008fission, ramos2018fission} that occupy a common area but are often divided into subgroups that frequently come together and split from each other. In species with a high degree of these dynamics, the entire group is never found together, and  subgroups of varying sizes spread over large areas, travelling independently of other subgroups \cite{ramos2011no} but also coming together often \cite{pinacho2017influence}. In terms of group-decision making, fission-fusion dynamics can be seen as the result of a lack of consensus between individuals who may have different priorities on their activities \cite{sueur2011collective}. However, in heterogenous, dynamic environments, this lack of consensus may have advantages in terms of distributed foraging and information gathering: individuals may forage in different regions of the group home range, while in subgroups with different combinations of individuals \cite{ramos2018quantifying}. Thus, a subgroup will often contain individuals who have been foraging in (and perhaps have knowledge about) different regions of their home range.

Spider monkeys (\textit{Ateles} spp) are frugivorous primates living in tropical forests, where fruit can be found in ephemeral patches that vary in time and space depending on the tree species \cite{difiore2008diets}. They have a high degree of fission-fusion dynamics \cite{ramos2014unraveling} in terms of the variation in subgroup size \cite{pinacho2017influence} and composition \cite{ramos2018quantifying}, and they occupy partially overlapping home ranges \cite{smith2016seasonal}. In previous work \cite{palacios2019uncovering}, we showed that spider monkeys share information about newly discovered fruiting trees. Focusing our observations on relevant trees during their whole fruiting period, we showed that individuals that were naïve about the presence of fruit in the focal tree tended to arrive with other group members that already knew about it. Using a random arrival null model, we also showed that the group as a whole finds out about the fruiting tree in fewer visits than if each individual had to find it on its own \cite{palacios2019uncovering}. 

Partly inspired by those results, Falcón-Cortés et al. \cite{falcon2019collective} used an agent-based model to explore the effects of information sharing and memory of visited patches on the collective estimation of food patch quality. When individual agents remember previously visited patches, move toward them depending on their quality, and copy others' visiting patterns if not finding food on their own, the group as a whole is able to compute the relative rank of food patches in the environment, concentrating its activity in proportion to the size of patches and focusing foraging efforts around the largest patches  \cite{falcon2019collective}. These studies show that sharing complementary information about heterogenous environments can effectively lead to an improvement in the group's overall knowledge, and thus could be considered a form of collective information processing, \textit{sensu} \cite{moussaid2009collective}.

Here we assume that information sharing takes place between individuals, in the context of a fluid grouping pattern where knowledgeable and naïve individuals coincide temporarily in the same subgroups and locations, being able to share information with others about areas they uniquely know. We also assume that an individual's core range represents, for a given season, the area where it knows the location of available fruiting trees relatively well. This assumption is supported by the species' wide daily travel distance (between 500 and 4500m with means between 1750 and 3311m in different study sites \cite{wallace2008factors}) when an individual visits around 10 different fruiting trees \cite{suarez2006diet}. Individual core ranges vary across seasons as would be predicted based on fruit availability, with dry seasons showing larger and less overlapping core ranges than wet seasons \cite{smith2016seasonal}.

\begin{figure}[H]
    \centering
    \includegraphics[width=1\textwidth]{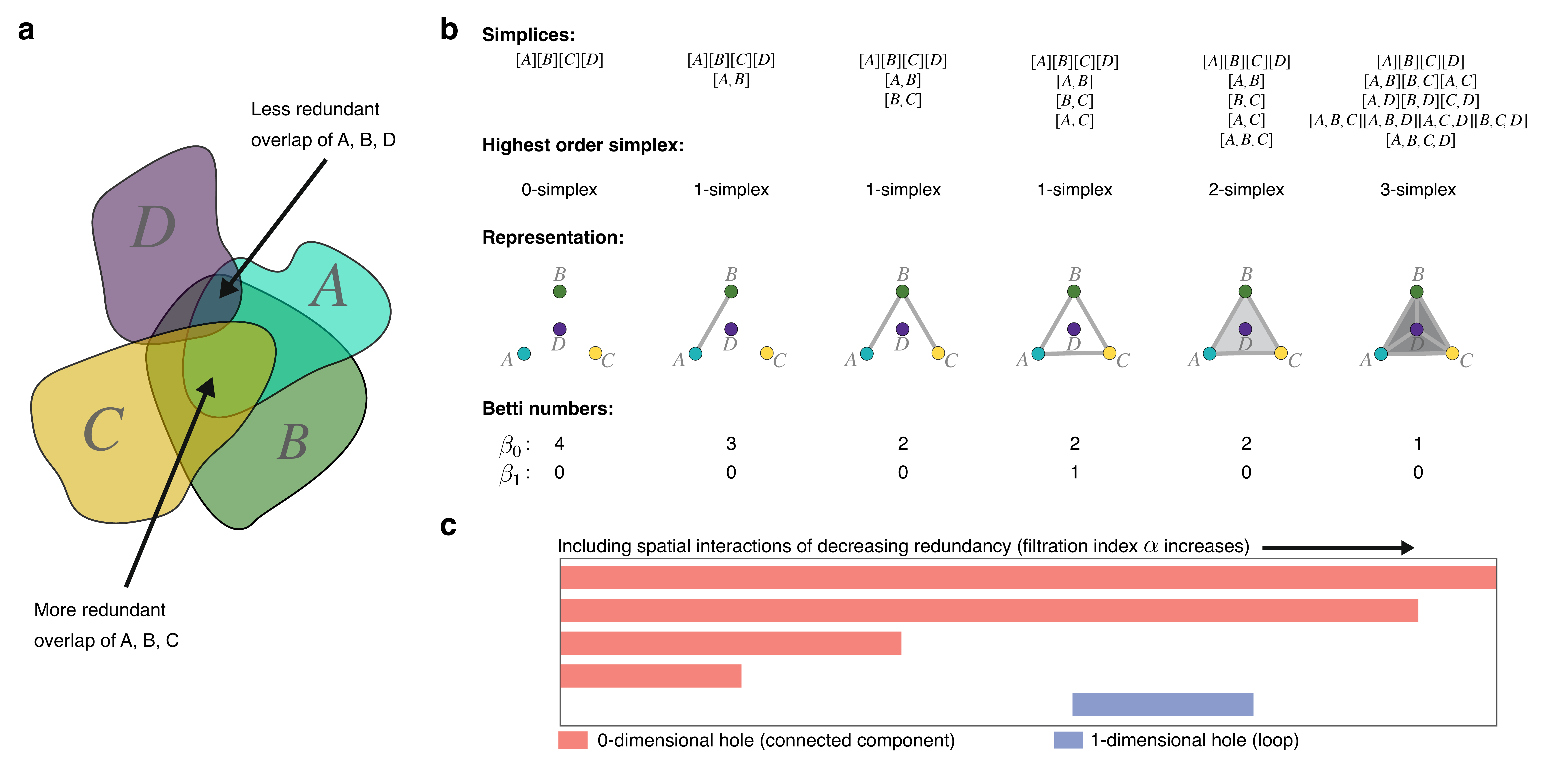}
    \caption{Strategy to analyse spatial overlaps in individual core ranges as higher-order interactions in a simplicial complex.  (a) Four hypothetical areas from individuals $A$, $B$, $C$ and $D$ overlap to different extents. The balance of a spatial interaction between a set of any size, $n$, is defined by the intersection/union ratio $w$, which in our idealised model, scales with $n$ as $w^*=1/(n+1)$ (see Methods for details). We use a filtration index, $\alpha$, which describes how far from $w^*$ a given $w$ is, to filter the sets of individual core ranges, from those with highly redundant spatial interactions (low $\alpha$) to those with less redundant ones (higher $\alpha$), as outlined in detail in the Methods. In this example, the areas $A$, $B$ and $C$ have a more redundant spatial interaction than the areas $A$, $B$ and $D$ due to their higher intersection/union ratio. Thus, the 2-simplex $[A,B,C]$ is created at lower $\alpha$ values than the 2-simplex $[A,B,D]$. (b) Different configurations of the resulting simplicial complex as $\alpha$ increases (going from left to right, adding links depending on the redundancy of the spatial interaction). Assuming no interactions would represent these four areas as nodes in an unconnected graph. As we incorporate spatial interactions according to a decreasing intersection/union ratio, we introduce links that join the initially disconnected nodes. First the link between $A$ and $B$, who intersect the most, then $B$ and $C$, later $A$ and $C$. As these simplices appear and close a circuit, they leave a hole of dimension 1 which is filled by a 2-simplex when our filtering includes the interaction between the three of them. Finally, when incorporating the smaller and less redundant spatial interaction between $D$ and the rest, we obtain a 3-dimensional simplicial complex which contains all possible simplices within it, including the three-dimensional space inside the tetrahedron, representing the four-way interaction between all nodes. Above each simplicial complex we note the simplices it contains (in brackets), and the order of the resulting simplicial complex, given by the highest-order simplex it contains. Below each simplicial complex are its Betti numbers $\beta_0$ and $\beta_1$. (c) The corresponding barcode diagram representing the persistence of the four connected components (red bars) as we incorporate less redundant and more unique interactions and the persistence of a 1-dimensional hole (blue bar). The Betti numbers at each step in the filtration correspond to the number of red ($\beta_0$) and blue ($\beta_1$) bars present from that step onward along the filtration.
    }
    \label{fig:strategy}
\end{figure}

The way in which individual core ranges overlap shows that, normally, at least two individuals (but not the whole group) coincide in most portions of a group's range \cite{smith2016seasonal}, such that range overlaps can be thought of as higher-order interactions between two or more individuals. Therefore, typical network approaches to describe the full space-sharing structure of the group are inadequate due to their inability to capture multibody (non-dyadic) interactions between individuals \cite{Silk2022, silk2023conceptual}. \textit{Simplicial sets} are increasingly applied to model such scenarios where non-dyadic interactions are key features of network structure   \cite{Iacopini2024, Iacopini2024Temporal, hasenjager2024social, Lin2024}. Simplicial sets are generalisations of ordinary networks which allow for arbitrarily many nodes or individuals to be connected by a single higher-order edge called a \textit{simplex} (Figure \ref{fig:strategy}). For spatial problems, a particularly relevant class of simplicial structure is the \textit{simplicial complex} \cite{Silk2022} which imposes the additional assumption of \textit{downward closure}; meaning that the simplicial set must contain all possible nested lower-order simplices of any simplex \cite{silk2023conceptual, Serrano2020, Otter2017}. For example, a simplicial complex that includes a triadic connection between three individuals must also contain the three dyadic connections between these individuals. This assumption of downward closure provides a suite of additional computational tools for studying higher-order structure \cite{Otter2017}.

Simplicial complex structure is often described through the computation of \textit{Betti numbers}. There is a Betti number for each dimension, denoted $\beta_i$, which gives the number of holes of a particular simplicial complex in the $i$-th dimension. For instance, $\beta_0$ is the number of connected components, $\beta_1$ the number of loops and $\beta_2$ the number of voids \cite{Otter2017}. For higher dimensions, we cannot visualise the holes, so the Betti numbers are especially useful to describe higher dimensional topological structure. In the context of a higher-order network structure of spatial overlaps, the topological holes described by the Betti numbers might result from individuals, or subsets of them, using areas uniquely with respect to others while maintaining some degree of redundancy in the common intersection. Thus, Betti numbers may characterise complementarity in the information possessed by different subsets (represented here by simplices). In addition, simplicial centrality measures \cite{Serrano2020} allow the identification of simplices that are well-connected to others in the simplicial complex.

We hypothesise that, by pooling complementary information about the location of feeding trees, the group as a whole acquires a more complete knowledge of a complex, dynamic environment during a given season than any individual on its own. In particular, given our knowledge of spider monkey socio-spatial ecology, we expect to find evidence of complementarity in the structure of the simplicial complex representing the overlaps of the individual core ranges. Our strategy, outlined in Figure \ref{fig:strategy}, involves first calculating individual core ranges and their relative intersections. We assume that complementary foraging information can be captured through the intersection/union ratio for any given set of individual core areas, representing a balance between redundant and uniquely known information. The intersection is where the set can coincide to share information about those areas known uniquely by subsets of individuals in the rest of their union.

We use varying thresholds of this intersection/union ratio to connect sets of individuals with decreasingly redundant overlaps into simplicial complexes. Then, we explore the structure of these complexes, searching for holes in various dimensions (given by the Betti numbers) and their persistence for different thresholds of redundancy. We interpret these holes and their persistence as evidence of complementarity in the foraging information sharing by the whole network. We also compare the simplicial complexes between the dry and wet seasons, as the uncertainty about the foraging environment may be greater in the former. We expected to find evidence for more complementarity in the information being shared during the dry seasons. Finally, we analyse the centrality patterns in the structure of the simplicial complex, to identify those sets of individuals that are key in terms of sharing foraging information with others by participating in more simplices.

\section*{Results}
\subsubsection*{Individual core ranges overlap only partially}
\begin{figure}[h!]
    \centering
\includegraphics[width=\linewidth, trim={40 260 0 100}, clip]
{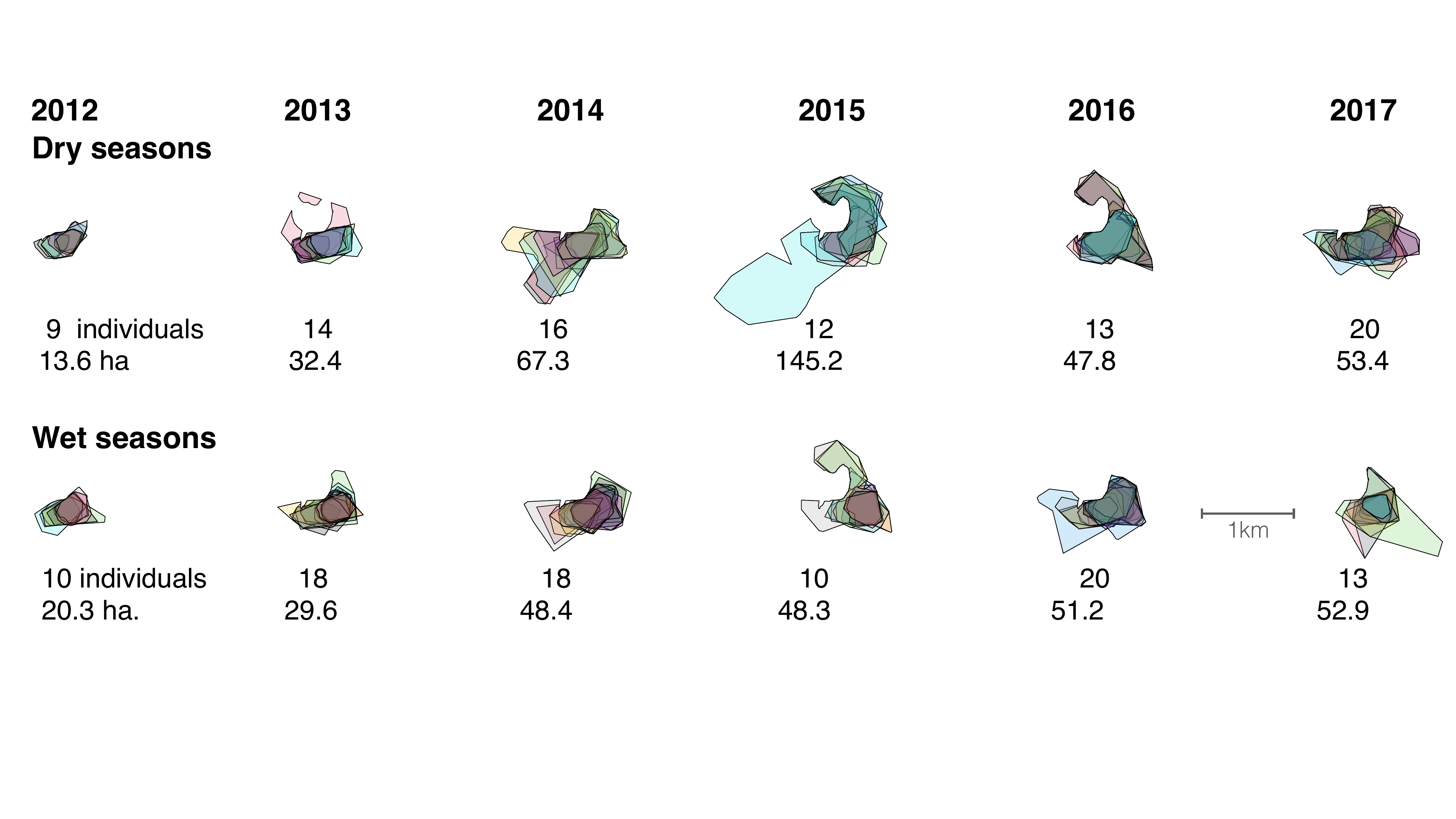}
    \caption{Maps of overlaid individual core ranges, for different years and seasons (top row: dry seasons; bottom row: wet seasons). Each polygon filled with a different colour shade corresponds to an individual (scale bar in the lower right). Below each map is the number of individual core ranges overlaid, as well as the area comprised by their union, in hectares. The northwestern part of the group's home range borders the Punta Laguna lake (see \cite{smith2016seasonal} for more information about the spatial ecology of spider monkeys at the study site).}
    \label{fig:maps_core_areas}
\end{figure}

A visual inspection of the overlaid individual core ranges during a given season shows that different subsets of individuals predominantly use different parts of a group's home range (Figure \ref{fig:maps_core_areas}). The area comprised by the union of the individual core ranges was between 20 and 70 ha throughout the study, with the exception of the dry season of 2015, when one individual occupied a core range that was much larger than anyone else's (bringing the union to 145.2 ha). There was no statistically significant increase in the area of the union with the number of individuals overlaid in each season ($\rho_{10}$ = 0.06, $p= 0.85$) nor a significant difference in the union area between seasons (ANOVA $F_{1,10}$ = 0.875, $p=0.37$).

\begin{figure}[H]
    \centering
    \includegraphics[trim={0 0 0 0}, clip, width=1\linewidth]{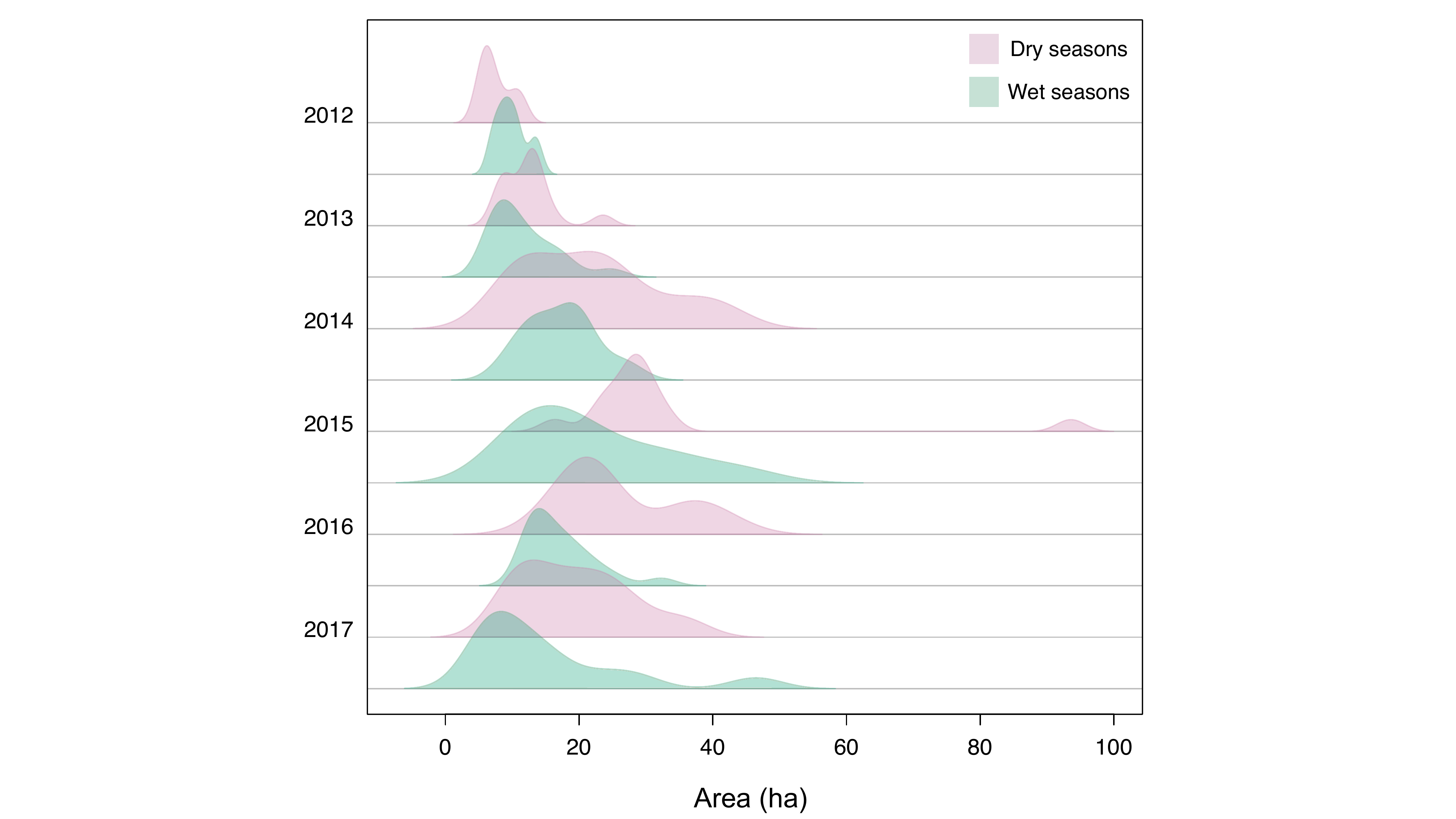}
    \caption{Density plot of individual core ranges, split by year and season (magenta: dry seasons; green: wet seasons). Bandwidth for the density plots was chosen via the \textit{nrd0} option in the density function in $R$, which adjusts for standard deviation and interquartile range.}    \label{fig:CA_density_plot}
\end{figure}

Figure \ref{fig:CA_density_plot} shows the density distribution of individual core ranges for the different years and seasons. Individual ranges differed significantly between years and seasons (ANOVA interaction between years and seasons, $F_{1,135}$=1.94, $p$=0.09; year: $p$<0.001; season: $p$ <0.001). During a given year, individual spider monkeys occupied larger core ranges during the dry season than during the wet season. This is particularly evident in 2014, 2015 and 2016.

\subsubsection*{The extent of overlap varies predictably with the number of individual core ranges}
We developed a theoretical prediction of the extent to which the core ranges of a set of individuals of size $n$ should intersect in order to strike a balance between the opportunities to coincide in the common overlap and the amount of unique knowledge each member of the set should hold outside of that area. In order to develop this prediction (detailed in the Supplementary Information), we first assume that a particular spatial overlap structure in a given subset of individual core ranges has a degree of complementarity depending on the amount of information that can be transferred between individuals. This information transfer, in turn, is proportional to the area known uniquely by some individuals but not others, and to the opportunities they have to coincide in a common area in order to share their uniquely known information. The higher the transfer of information, the more complementary their relative use of areas is. We also assume that individuals are distributed uniformly within their own areas, and independently from each other. Then we constructed an objective function describing this transfer of information and optimised it under the additional assumption of an homogeneous foraging ability in the population, i.e., individual core ranges of identical sizes (see Supplementary Information for more details). We found that the optimal intersection/union ratio for a set of $n$-many individuals is $w^*=1/(n+1)$. 

\begin{figure}[H]
    \centering
    %left, bottom, right, top
    \includegraphics[trim={10 150 0 150}, clip, width=1.015\textwidth]{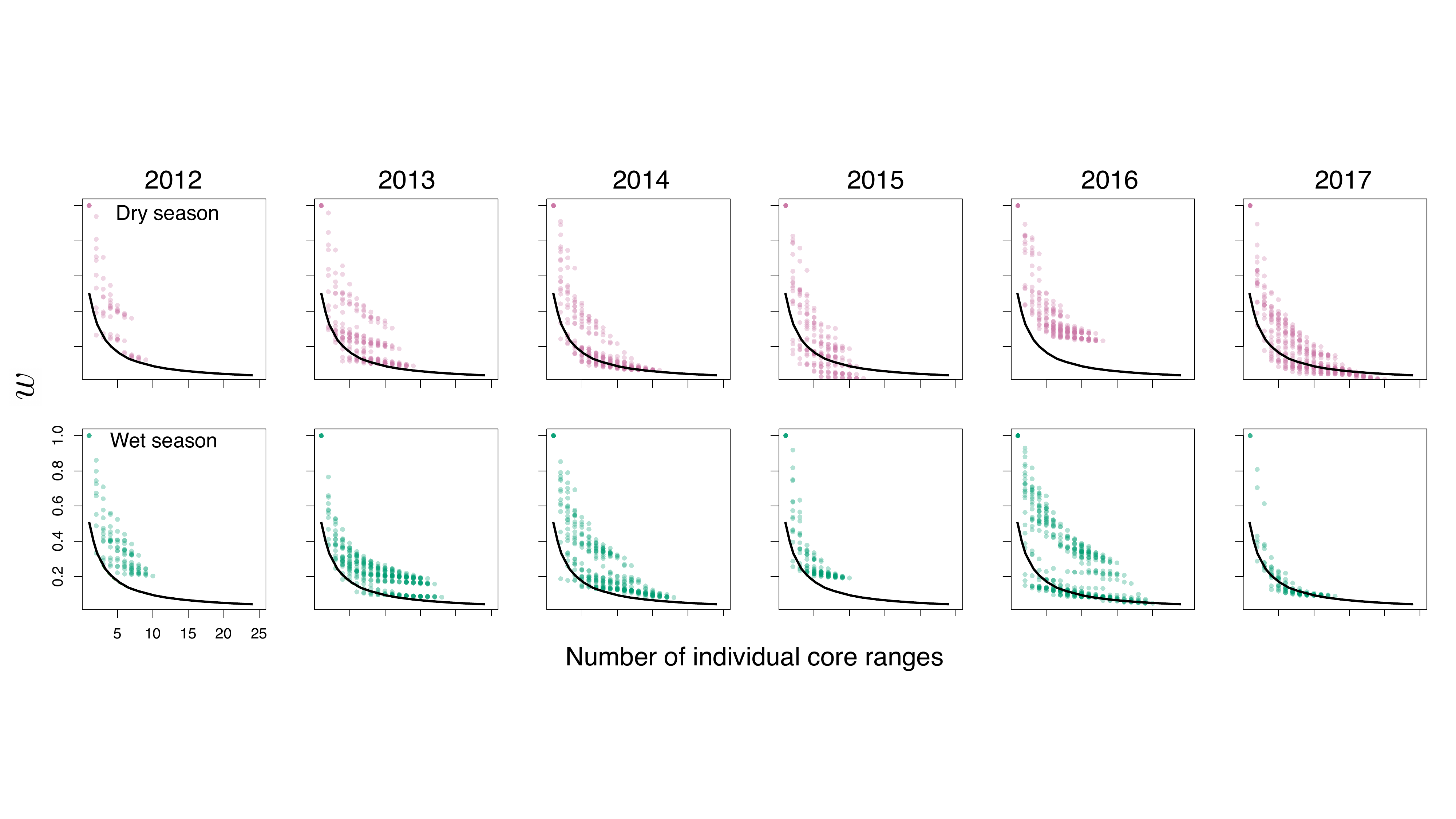}
    \caption{Scaling relationship between the number of individual core ranges and its intersection/union ratio, $w$. Each dot corresponds to a particular set of a given number of individuals (see Methods for details). The solid black line corresponds to the scaling relationship $w^*_n=1/(n+1)$. Top row: dry seasons; bottom row: wet seasons.}
    \label{fig:w_and_n}
\end{figure}

The above relationship predicts that larger sets of individual areas should have a lower intersection/union ratio, according to the convex decreasing curve shown in Figure \ref{fig:w_and_n}. The empirical values of this ratio similarly show a convex decreasing trend for all overlapping sets, in many cases qualitatively fitting the theoretical prediction (Figure \ref{fig:w_and_n}). Overall, smaller sets of up to five individuals vary more widely in the value of $w$ than larger sets. The qualitative fit between the optimal prediction and the observations is particularly good for some seasons (dry seasons in 2014 and 2017 and wet seasons in 2013, 2016 and 2017), but less so in others (dry season in 2016, wet season in 2015). At fixed $n$, there also appears to be a multimodal distribution of the observed values of $w$, with the lowest values resembling the theoretical prediction and others being higher than predicted, suggesting other processes that are not captured by our theoretical model. In sum, the prediction of an optimal value of $w*$, representing a complementary balance between redundant and unique information, is overall consistent with the observed patterns, but overlaps tend to be larger (i.e. more redundant) than predicted by our model alone.

The observed values of $w$ did not vary significantly between seasons but they varied significantly between years (ANOVA interaction between years and seasons, $F_{1,3084}=9.56$, $p<0.001$; year: $p<0.001$; season: $p=0.24$).

\subsubsection*{The structure of simplicial complexes shows holes at various dimensions}

The previous results on the relevance of $w^*$ led us to use it as a reference for constructing simplicial complexes depending on how far a given set of individual core ranges was to the predicted $w^*$. Each simplicial complex has an $\alpha$ value calculated from their value of $w$, taking into account how the theoretical optimal prediction $w^*$ varies with number of individuals, $n$. At low values of $\alpha$, simplicial complexes are constructed only containing sets of individual areas that overlap considerably (i.e. have highly redundant spatial interactions), while high values of $\alpha$ imply the additional inclusion of sets of individuals that overlap less (i.e. have relatively more unique spatial interactions). 

\begin{figure}[H]
    \centering
    %left, bottom, right, top
    \includegraphics[width=1\textwidth] {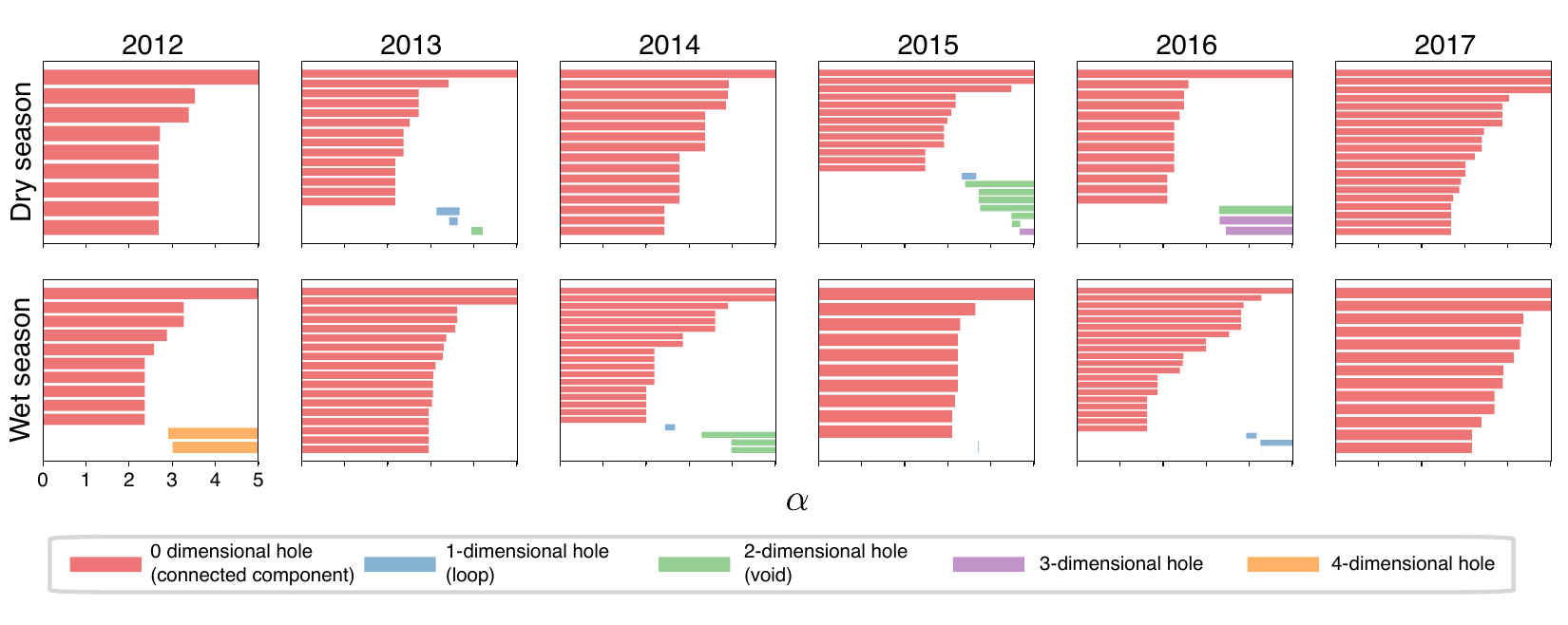}
    \caption{Persistence barcodes for the simplicial complexes in each season, including the simplices of individual core ranges formed by decreasingly redundant interactions, as $\alpha$ increases. Top row: dry seasons; bottom row: wet seasons. Each barcode shows, for increasing values of $\alpha$, the persistence of holes of different dimension as specified in the legend. The number of holes of each dimension $i$ at a given filtration step corresponds to each Betti number ($\beta_i$). The beginning and end of each bar correspond to the values of $\alpha$ at which the hole appears and disappears, respectively. By definition, at $\alpha=0$, each individual in the sample is a connected component on its own. While one connected component will persist for all values of $\alpha$, others which are initially present become joined with others and disappear as $\alpha$ increases and less redundant interactions are used to join subsets of individual core ranges into simplices (see also Figure \ref{fig:strategy} for an explanation of the filtration and the resulting persistence barcodes). Note that our persistence barcodes do not finish with trivial structures (containing each possible simplex) due to the inclusion of a maximal simplex size, $n_\text{max}$.}
    \label{fig:barcodes}
\end{figure}

When we use $\alpha$ as a filtration parameter, including in the simplicial complex only those simplices that are below the $\alpha$ threshold, we find evidence of structural holes at various dimensions (Figure \ref{fig:barcodes}).  Multiple connected components, representing holes in the lowest dimension, persist for values higher than $\alpha=2$, especially in the dry seasons. In some circumstances, only when including the overlaps with the lowest level of redundancy do the simplicial complexes become fully connected. The dry season of 2017, in particular, shows 3 connected components for all values of $\alpha$. Bi-dimensional holes ($\beta_1$) and voids ($\beta_2$) are common when including interactions beyond $\alpha=3$ and often persist for several units of $\alpha$. Holes of dimension 4 ($\beta_3$) only appear in the dry seasons of 2015 and 2016, while two long-persisting holes of dimension 5 ($\beta_4$) appear in the wet season of 2012. In the dry season of 2015,  holes of four different dimensions ($\beta_0$ to $\beta_3$) appear at some $\alpha$ values, representing a particularly rich spatial structure across the filtration. 

If we assume that a hole of any dimension in the simplicial complex is the result of unique areas being used by some individuals in the group but not by others, then we can interpret holes and their persistence over different values of the filtration parameter as evidence of complementarity in the foraging information shared by these subsets. The existence of these holes implies that pockets of unique knowledge may exist among some sets of individuals even as we include spatial overlaps with less redundant information (higher $\alpha$). Overall, these results provide compelling evidence that there are many areas occupied uniquely by subsets of individuals of various sizes, and, if we interpret these patterns of overlaps as indications of partial information sharing, there is potentially a large amount of complementary information being shared.

A qualitative exploration of the persistence barcodes in Figure \ref{fig:barcodes} suggests that the dry season contained more topological holes, which were also more persistent, providing stronger evidence for knowledge complementarity in the dry seasons. To quantify the patterns in each persistence barcode, we developed a \textit{filtration complementarity index}, which measures the interval over $\alpha$ for which different holes last over the filtration, and gives more weight to those of higher dimension. The values of this index show no clear evidence for simplicial complexes from the dry seasons to have higher complementarity than for those of the wet seasons (ANOVA $F_{1}$=2.2, $p=$0.17; Figure SI4), although our power to detect an effect is weak given our small sample size of 6 years. The pattern is similar for the correlation with the variation in fruit abundance across fortnights for each season. While seasons with higher variation in fruit abundance have slightly larger values of complementarity on average (Figure SI1), we did not find statistical support for this pattern ($\rho$ = 0.42 and $p$=0.16).

\subsubsection*{Simplicial centrality increases with size of the simplex}
To examine the structure of the simplicial complexes, we utilise the maximal simplicial degree centrality, a generalisation of degree centrality in dyadic networks \cite{Serrano2020}. Across the majority of seasons, we found a clear sigmoidal relationship between the centrality of a simplicial complex and its size. Maximal degree centrality increases gradually with size of the simplices and then more steeply, levelling off or decreasing after the simplex is of size 5. This generally implies that simplices containing more individuals are involved in a larger and more variable number of spatial interactions with other simplices, while above a certain size there is a plateau in this effect. We found no clear trends in the relationship between the simplicial degree centrality of a simplex and the average age, the proportion of males or the proportion of immigrant individuals in the individuals comprising it (see Supplementary Information, Figures SI2-4). Note that the increasing trend of centrality with simplex size is consistent when varying the parameter $n_\text{max}$ (see Supplementary Information, Figures SI5-9), implying that it is indeed larger simplices which are more central and not those of some intermediate size which we could not observe due to our choice of $n_\text{max} = 6$.

\begin{figure}[H]
    \centering
    %left, bottom, right, top
    \includegraphics[width=\textwidth]{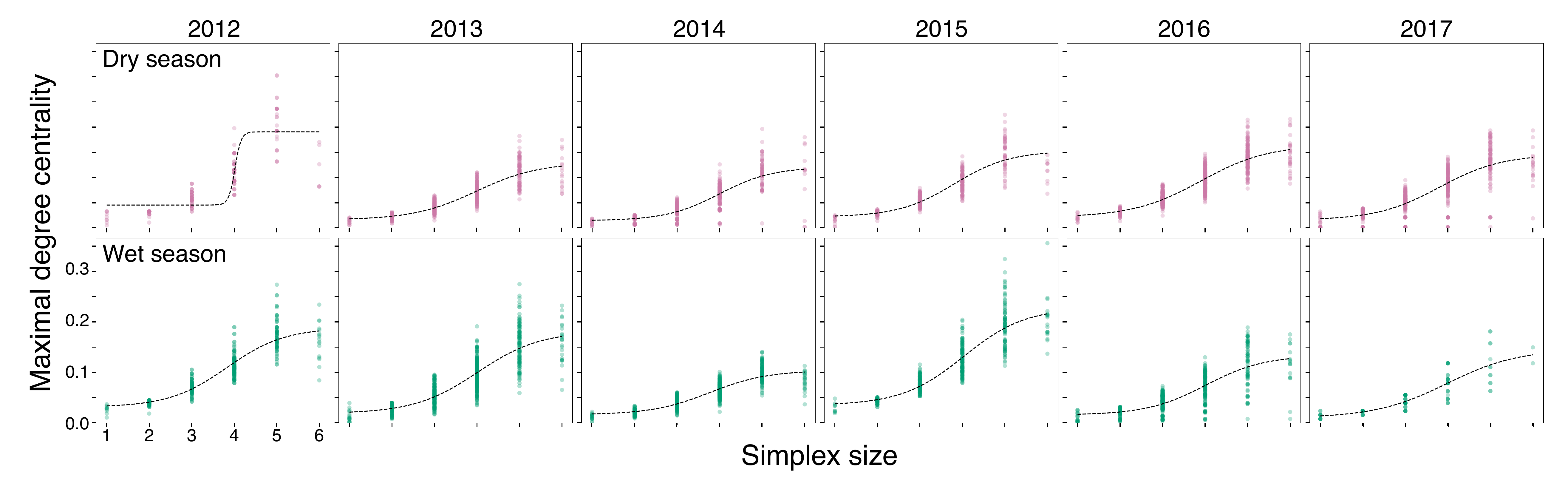}
    \caption{Maximal simplicial degree centrality values for simplices of different size. Each dot corresponds to a simplex formed by the intersection of individual core ranges assuming $\alpha \leq 4$. The dashed black line corresponds to a continuous sigmoid function fitted to the centrality values using the non-linear least squares method from the \textit{Scipy} package in Python \cite{SciPy2020}. Top row: dry seasons; bottom row: wet seasons.}     \label{fig:centrality_and_set_size}
\end{figure}

\section*{Discussion}\label{discussion}

We have studied the higher-order spatial interactions between individual spider monkeys, who live in a highly variable foraging environment and have a high degree of fission-fusion dynamics. The majority of a group's range is covered by different combinations of, quite often more than two, individual core ranges. Simplicial complexes allowed us effectively represent these higher-order interactions, providing a methodology to quantify the complementarity of different subsets of individuals of varying size. These simplicial complexes contained holes that persisted for various values of the redundancy of interactions. These pockets of unique knowledge among some sets of individuals remained even when including less redundant interactions than predicted by $w^*$. In other words, the spatial interaction networks did not become fully connected until we included very weakly connected sets of individuals, which are likely sharing a large amount  of unique knowledge even if coinciding in small intersections. Overall, our results provide strong support for our prediction of complementary patterns of information sharing.

In agreement with our prediction of the optimal spatial arrangement for maximum information transfer, spider monkeys seem to be balancing the intersection/union ratio of their core areas, or the extent of redundant and unique knowledge that an  interaction would provide to a given set. Optimality approaches to collective behaviour are scarce, due to the difficulty in identifying the currency and level of optimization \cite{marshall2024aims}. Here we have informed our representation of higher-order interactions as simplicial complexes with this theoretical optimal prediction of a convex decreasing function of the size of the subset. This optimality argument, developed in detail in the Supplementary Information, assumes that the intersections between individuals comprising each simplicial complex provide an opportunity to share unique information that each one (or subsets of them) possess. In other words, our prediction is simply based upon the balance between redundancy and uniqueness in the use of space by any given set of individuals. This balance is not what one would propose for a central place forager (e.g. hooded crows, \textit{Corvus corone cornix} \cite{sonerud2001ignorant}) which would only share a small portion of redundant areas for sets of any size, as the opportunities for information sharing would only take place in a small, regularly used central area within their range (e.g. their roost). Our optimality argument also rests on the assumption that each group member benefits from information pooling in a way that surpasses the costs of feeding competition \cite{clark1984foraging} and the alternative benefit of gathering information individually \cite{valone2002public}. 

While other, alternative models could possibly explain the observed patterns, ours simply assumes a balance between the opportunities to coincide in overlapping areas and the maximum amount of knowledge that can be shared about areas outside of those overlaps, applicable to any number of individual areas. For example, the most redundant interactions could be due to individuals cohesively exploring common areas for other, social reasons. Conversely, values below the predicted line are rare: sets of individual ranges do not seem less redundant than would be predicted based on our optimality argument. This is consistent with previous findings on the same species \cite{ramos2020collective} indicating that repulsion between individuals is less relevant than attraction as a mechanism behind fission-fusion dynamics.

Particularly for smaller subsets, between 2 and 6 individuals, the value of the intersection/union ratio falls steeply with $n$, both for the optimal $w^*$ and the observed values of $w$. This suggests that the balance between redundant and unique knowledge at this range of subset sizes is more sensitive to the addition of new areas, contrary to the situation with subsets of 10 individuals or larger. It is interesting that we found a similar pattern in the simplicial centrality analysis: this metric increases sigmoidally as simplices increase in size, with a maximum rate of increase in simplices of intermediate size (4 to 6 individuals) and then levelling off at larger simplex sizes of 7-9, depending on the season and the maximum simplex size considered (Figures SI5-9). From the overall patterns of simplicial centrality, we can speculate that simplices representing subsets of intermediate size could be key in maintaining updated levels of foraging information, as their centrality is more sensitive to simplex size and also have their intersection/union ratio poised at the most sensitive range with respect to the addition of new individuals (and are therefore the most dynamic ones in terms of the redundant and unique knowledge they possess).

Sueur et al. \cite{sueur2011collective} predicted that broker individuals, who connect different clusters of the association network, are particularly important in fluid groups with high degrees of fission-fusion dynamics. Our simplicial centrality results show that subsets of increasingly many individuals participate in increasingly many spatial interactions with other simplices. The fact that we did not find a clear relationship between simplicial centrality and the age and sex of the adults, or their recent immigration status, suggests that rather than being an attribute of individuals (as in killer whales, for example: \cite{brent2015ecological}), the asymmetry in foraging information is predominantly the result of spider monkeys' high mobility and fluid grouping patterns. This had already been suggested by previous work on the factors that determine an individual's decision towards fission or fusion \cite{ramos2014unraveling} and is compatible with the idea that spider monkeys are collectively pooling their foraging information in a dynamic fashion \cite{palacios2019uncovering}. Of course, many other factors related to individual age and sex classes shape the structure of dyadic association networks \cite{ramos2009association}, but here we are uncovering those aspects of fission-fusion dynamics that relate to foraging knowledge as a dynamic, distributed process which varies depending on all adult members' knowledge about a dynamic environment. Future work analysing associations in subgroups as higher-order interactions will be useful to evaluate the relevance of foraging knowledge as we have uncovered it here as an additional influence on the structure of association networks.

We did not find stronger evidence for complementarity in the dry compared to the wet seasons. While it appeared as though the simplicial complexes had more holes of various dimensions during the dry seasons, we were unable to detect statistically significant differences using a composite index of filtration complementarity. During the dry seasons, spider monkeys tend to rely on less abundant tree species and thus exploit a more heterogenous environment. During the wet season, the hyperabundance of \textit{Brosimum alicastrum} \cite{smith2016seasonal} would make it less valuable to share information about scarce sources of food. However, we did not find significant differences in the temporal variation of the biweekly fruit abundance between the seasons. While our sampled number of seasons is small, it is possible that rather than temporal variation, it is the spatial variation that is higher in the dry than in the wet seasons. During the dry seasons, fruit is more scarce and dispersed, subgroups tend to be smaller and home ranges larger \cite{smith2016seasonal}. It would be interesting to test the information content of different regions of the home range, using more accurate tree distribution and abundance data than was available for this study.

More generally, finding resources in a heterogenous, dynamic environment can be seen as a problem of pattern recognition, which can be solved effectively by systems with some degree of distributed information processing. This capacity can be found in various natural systems, recently described as `liquid brains' \cite{sole2019liquid}, ranging from social insect colonies to the immune system. These systems process information via interactions between mobile agents that interact only temporarily when they happen to be within reach. As an example, Vining et al. \cite{vining2019mobility} modelled a set of simple agents that could only communicate with others within a short radius but were able to move about, thus increasing the number of agents with whom they could exchange information over time. This system was able to solve a global problem of density classification (converging to the binary state that was more common in the beginning of the run amongst all agents). Crucially, agents were able to solve the most difficult versions of the problem (where the difference between the majority and minority binary state was very small) only when moving at intermediate speeds. This allowed them to balance the frequency of interactions with others in their current neighbourhood and those in other locations. In other words, an optimal degree of mixing between agents led the group to solve more difficult problems effectively \cite{vining2019mobility}. This optimal degree of mixing is analogous to the optimization of $w$ we used here: an optimal area of intersection between individual use areas allows monkeys to coincide to share information while exploring enough area on their own which can be shared. Unique information, gathered by individual members of the group or subsets of them, becomes complementary when monkeys coincide in redundant areas and can share what they uniquely know. The redundancy/uniqueness balance is also more generally related to that between exploitation, where more redundant information is gathered about the environment, and exploration, where more unique information is gathered \cite{hills2015exploration, luppi2024information}. 

Collective information processing can result from individual members sharing complementary information, as we have found here, or it may also include \textit{synergistic} information \cite{williams2010nonnegative}, whereby two or more individual estimations of a target variable not only are complementary in their estimation but combine their information in such a way as to produce new knowledge \cite{bettencourt2009rules}. An example would be if one subset of individuals would contribute the location of a food source and another subset the timing of the fruiting of that source. The latter could be done by individuals who happened to visit the same trees repeatedly, gathering unique information about their fruiting status (as  has been observed in other high fission-fusion species like the chimpanzee (\textit{Pan troglodytes};   \cite{janmaat2016spatio}). The resulting, combined knowledge by both subsets of individuals would be synergistic in the sense of allowing all of them to exploit the food source according to its location and timing. Further empirical and modelling work could explore these ideas and search for synergy in collective information processing about the foraging environment in this and other species.

Using a theoretical prediction of the optimal balance between the use of redundant and unique areas, as well as simplicial complexes to represent higher-order interactions, we have been able to show complementarity in information sharing, a hallmark of collective information processing \cite{bettencourt2009rules, garg2025causal}. This work exemplifies the usefulness of higher-order methods, as we have been able to uncover details in the spatial structure which could not be found in a dyadic representation of the same structure. While these methods come with greater complexity over the dyadic approach, they are extremely well suited to tackling problems surrounding collective behaviour. Spider monkeys with different knowledge of their foraging area contribute with different pieces of information to group-level knowledge, in such a way that the group as a whole can access more foraging trees in a dynamic landscape than any individual could on its own. This complementary information sharing in higher-order spatial networks would be a compelling example of collective intelligence in natural conditions \cite{hutchins1995cognition}.

\section*{Methods}\label{methods}
\subsection*{Data collection}
Ranging and subgroup composition data was collected by experienced observers between January 2012 and December 2017 in a habituated group of Geoffroy's spider monkeys (\textit{Ateles geoffroyi}) living around the Punta Laguna lake in the \textit{Otoch Ma'ax yetel Kooh} protected area, in the Yucatan peninsula, Mexico. This group was the subject of a long-term, continuous study from 1997 to 2020 \cite{ramos2018ecologia}. We defined dry seasons from November to April and wet seasons from May to October \cite{smith2016seasonal}. The spider monkeys' foraging environment differs crucially between the two seasons, due mainly to a very abundant species (\textit{Brosimum alicastrum}) fruiting during the wet season \cite{smith2016seasonal}. Therefore, fruit during the wet season is more abundant and homogeneously distributed, while during the dry season spider monkeys use more species that are less abundant overall and thus face a more heterogenous and uncertain environment \cite{smith2016seasonal}. 

We tracked changes in fruit abundance by recording, every two weeks, the number of trees with fruit for each species in a phenological trail. This trail contained a sample of 10 individual trees from 16 species, including 12 which comprise > 50\% of the monkey's diet \cite{ramos2001patterns, pinacho2010patrones}. We estimated the density of each of these species using data collected in eight 0.25ha plots (total of 2ha) and fourty-eight 100m x 2m line transects (total 0.96 ha) within the home range of the spider monkeys (see \cite{bonilla2008forest} for more details about the vegetation sampling).

During 4-8 hour-long daily subgroup follows, we performed subgroup scans every 20 minutes, noting the subgroup composition and its position using a hand-held  global positioning system located roughly in the center of the  subgroup (subgroups can be dispersed over a maximum diameter of 50m). Individual core range estimation was based on all the subgroup locations where a given individual was present. We restricted the analysis to adults (individuals >5 years by January 1st of each year analysed) given that juveniles and infants do not travel independently of their mother and thus would not make any independent contribution to the information pool of a given season. We also confined the analyses to adult individuals with a minimum of 100 scan sample locations per season. Table SI1 contains the number of samples and observation days for each study individual that fulfilled these criteria in each season and year.

\subsection*{Data analysis}
\subsubsection*{Individual core range estimation}
We used individual locations for the estimation of individual core ranges, defined as the 60\% utilization distribution generated with the adaptive Local Convex Hull (a-LoCoH) method developed by \cite{getz2007locoh}, using the T-LoCoH package \cite{lyons2013home} for the R programming environment \cite{getz2007locoh} with a constant value of 15000 for the \textit{a} parameter for hull construction across seasons. The use of the 60\% utilization distribution as a meaningful area is validated by previous work where we showed that it lied at the inflection point in the curve of isopleth values versus proportion of home range covered \cite{ramos2013site}. Core ranges defined in this manner represent the area where individuals concentrated their regular movements, excluding more sporadically used areas. For more details on the calculation of individual core ranges, see \cite{smith2016seasonal}.

\subsubsection*{Individual core range overlap computation}
We calculated the intersections and unions between individual core ranges for each season using the \textit{Spatstat} package \cite{baddeley2015spatial} for the R programming environment \cite{rcoreteam2021R}. For a set $C = \{C_1,\dots, C_n\} $ of $n$ individual core ranges, we used the ratio denoted as $w\in[0,1]$ of the areas of their intersection and their union:
\begin{equation} \label{w_defined}
    w =\frac{\operatorname{Area}\left(\bigcap_{i=1}^n C_i\right)}{\operatorname{Area}\left(\bigcup_{i=1}^n C_i\right)}.
\end{equation}

\subsubsection*{Fruit abundance calculation}
We used the fruit abundance data from the phenological trail, together with the diameter at breast height of the trees monitored and the density for each species in the study site, to estimate fruit abundance and its variability in each season. We used the following formulation, based on \cite{smith2016seasonal}:
\begin{equation} \label{ifa}
    \operatorname{IFA}_{f} =\sum\limits_{i=\operatorname{1}}^{\operatorname{16}}{\operatorname{Tf}_i} \ {\operatorname{DBH}_i} \ {\operatorname{Den}_i}
\end{equation}
where $\operatorname{IFA}_f\geq 0$ is the index of fruit abundance for the monitoring period $f$ based on the 16 species monitored, $\operatorname{Tf}_i$ is the proportion of trees with fruit of species $i$, $\operatorname{DBH}_i$ is the sum of the diameter at breast height over all the sampled trees with fruit of species $i$ during period $f$ and $\operatorname{Den}_i$ is the tree density for species $i$ in the study site. We considered the $\operatorname{IFA}_f$ values for all the monitoring periods in a season to obtain the mean and the coefficient of variation of the $\operatorname{IFA}_f$ per season.

\subsection*{Higher-order network analysis}
\subsubsection*{Simplical complex construction}
For each dataset (i.e., each season), we represented the entire space-sharing network of the study group as a simplicial complex $K$. The nodes of the network were all the adult individuals in the group during the particular season of the dataset (Table SI1). Since all core ranges coincided in some small common area (\textit{e.g.} for sleep), we defined a maximal simplex size, $n_\text{max}$. This maximal size was needed because including interactions among more than $n_{\text{max}}$ individuals yielded a trivial simplicial complex structure (containing every possible simplex), and these larger interactions were less likely to be related to foraging activity, but rather to common sleeping areas in the centre of the group's home range. Then there exists a simplex connecting a subset of individuals if their corresponding intersection/union ratio, $w$, is non-zero, and if the size of the subset is no more than $n_\text{max}$. Upon observation, the value of $n_\text{max}=6$ was the largest which consistently resulted in non-trivial structures across all seasons. We computed the Betti numbers of the simplicial complexes for all seasons.

\subsubsection*{Filtration procedure}
We investigated how the final simplicial complex, $K$, was formed when adding simplices in decreasing order of redundancy. Between subsets of the same size, we assumed a higher relative spatial overlap, quantified by $w$, corresponded to a more redundant interaction. However, what constituted a redundant spatial overlap for information transfer depends upon the number of individuals in the subset, and therefore we required a scaling of the $w$ values with their corresponding subset size in order to make fair comparisons. To this end, we computed the $w$ value from the spatial overlap of $n$ individuals which optimised the information transfer when these individuals had equal domain knowledge and utilised their core ranges uniformly, balancing the size of the unique areas they could share information about with the opportunity to share them when coinciding with others. When this balance of redundant and unique information leads to an optimal information transfer, we define it as complementary. As detailed in the Supplementary Information, for $n$ individuals this optimal overlap was computed as: 
\begin{equation}
    w_n^* = \frac{1}{n+1}.
\end{equation}
This expression describes the amount of overlap required for a group of $n$ individuals to maximise their transfer of information under the given assumptions of uniformity and independence of movements. We then defined a parameter $\alpha$ to describe how far from $w^*$ a given overlap was. For the $i$-th non-zero spatial overlap in the data, consisting of $n(i)$ many individuals and with relative overlap value $w_i$, the corresponding $\alpha$ value was given as the solution to:
\begin{equation}
    w_i = f(\alpha)w_{n(i)}^* = \frac{f(\alpha)}{n(i)+1}
\end{equation}
where $f(\alpha)$ is some arbitrary positive and decreasing function. We picked $f(\alpha) = 5-\alpha$ as a linear function such that the corresponding $\alpha$ value for each set was non-zero, and each simplex has $\alpha>0$, thus ensuring we consistently begin our filtration with a trivial structure consisting only of 0-simplices. We varied $\alpha$ over the interval $[0, 5)$. For $\alpha\leq5$ each possible observation (including those with zero overlap, due to the inclusion of $\alpha=5$) is represented in the simplicial structure. Variation in the choice of the function $f$ would influence the axis of the persistence diagrams, but the same topological features would be observed. 

Then, we defined a sequence of simplicial complexes $K_0 \subseteq K_1 \subseteq \dots \subseteq K_N = K$. Here, $K_0$ is the set of individuals in the study group (without any simplices connecting them) and each $K_j$ is a simplicial complex with those simplices with $\alpha$ below the corresponding $j$-th lowest $\alpha$ value, i.e. corresponding to the intersection/union ratio $w$ above the corresponding $w$ threshold (possibly adding additional lower-order simplices to satisfy downward closure). This last procedure is justified on the basis of the clear tendency of overlaps to occur toward the centre of the group's range (see Figure \ref{fig:maps_core_areas}), which makes it unlikely that the areas of $n$ individuals will overlap significantly but the overlaps between subsets of $n$ will not. More generally, in higher-order spatial studies, it is standard to assume that if a higher-order interaction is significant, then all lower-order sub-interactions should be considered significant \cite{Silk2022}. This sequence is called a \textit{filtered simplicial complex}, and $\alpha$ is called the filtration parameter \cite{Otter2017}. Simplicial complexes which are lower in the sequence will consist of more redundant interactions in terms of their information sharing. We computed the Betti numbers for each simplicial complex in the sequence (the \textit{persistent homology}) for all available years and seasons in order to examine how the higher-order structure varied with the redundancy of interactions. Analysis was performed using the Gudhi package for Topological Data Analysis, implemented in Python \cite{gudhi2015urm}. 

\subsubsection*{Filtration complementarity index}
We quantified how much evidence of complementarity in information sharing there was in a filtered simplicial complex through the \textit{filtration complementarity index}. Structures with greater evidence of complementary sharing would be those with a larger number of higher order features (i.e. larger Betti numbers of high order), which persisted for longer throughout the filtration. Simplicial complexes with holes of higher dimensions may be described as more complex and present more evidence of complementary information sharing among a wider variety of simplices. We therefore defined the \textit{filtration complementarity index} for season $i$ as:

\begin{equation}
    \operatorname{FCI}_i = \left\langle \frac{B_i^{0}(\alpha)}{n_i} + \sum_{d=1}^{n} (d+1)B_i^{(d)}(\alpha) \right\rangle_{[0, 5)} 
\end{equation}
Here, $B_i^{(d)}(\alpha)$ is the $d$-th Betti number at the filtration value of $\alpha$ in the $i$-th season, $n_i$ is the number of individuals in the data in this season, and $\langle \cdot\rangle_{[0, 5)}$ is the average operator over the filtration interval $[0,5)$. We scaled the number of connected components by the number of individuals as to not give seasons with larger sample sizes a higher score automatically (as the number of connected components is highly dependent upon $n_i$). We did not scale the higher-order features as we did not expect them to be (directly) impacted by sample size differences observed in our data. Higher-order features were given greater importance through the scaling by $d+1$. We also tested other definitions of the filtration complementary index, including non-weighted or which did not consider the connected components. All of them yielded similar results.

\subsubsection*{Maximal simplicial degree centrality}

We additionally studied which subsets were the most central in the simplicial complex across years and seasons. For this, we implemented the maximal simplicial degree centrality measure as defined in \cite{Serrano2020}. This is a generalisation of the standard degree centrality used in dyadic networks to simplicial complexes, which identifies those subsets that appear in the largest number of simplices. This metric is designed so that the downward closure assumption does not produce biases towards higher or lower order simplices, which are not automatically classed as more or less central just for being part of a nested structure. We computed the centrality value for each simplex in the simplicial complex for each year and season at a filtration value of $\alpha=4$ (meaning we do not consider subsets with more unique interactions). Other values of $\alpha$ yielded similar results. We analysed how simplicial centrality varied with simplex characteristics (size, average age, proportion of males and proportion of immigrants). When analysing how simplicial centrality varies with simplex size, we considered variation in the parameter $n_\text{max}\in \{4,5,6,7,8,9\}$ to ensure our results were qualitatively robust to variations in the distribution of simplex sizes.

\section*{Code availability}
The underlying code for this study is available at
https://doi.org/10.5281/zenodo.15292113

\section*{Data availability}
The individual core ranges analysed, as well as the data used for constructing the simplicial complexes, are publicly available as part of the repository cited above.

\printbibliography

@article{brent2015ecological,
  title={Ecological knowledge, leadership, and the evolution of menopause in killer whales},
  author={Brent, Lauren JN and Franks, Daniel W and Foster, Emma A and Balcomb, Kenneth C and Cant, Michael A and Croft, Darren P},
  journal={Current Biology},
  volume={25},
  number={6},
  pages={746--750},
  year={2015},
  publisher={Elsevier}
}

@article{ramos2011no,
  title={No evidence of coordination between different subgroups in the fission--fusion society of spider monkeys (Ateles geoffroyi)},
  author={Ramos-Fern{\'a}ndez, Gabriel and Pinacho-Guendulain, Braulio and Miranda-P{\'e}rez, Ad{\'a}n and Boyer, Denis},
  journal={International Journal of Primatology},
  volume={32},
  pages={1367--1382},
  year={2011},
  publisher={Springer}
}

@article{mitchell2006complex,
  title={Complex systems: Network thinking},
  author={Mitchell, Melanie},
  journal={Artificial Intelligence},
  volume={170},
  number={18},
  pages={1194--1212},
  year={2006},
  publisher={Elsevier}
}

@incollection{difiore2008diets,
  title={Diets of wild spider monkeys},
  author={Di Fiore, Anthony and Link, Andres and Dew, J Lawrence},
   booktitle={Spider monkeys: Behavior, ecology and evolution of the genus Ateles},
  editor = {Campbell, Christina J},
  pages={220--235},
  year={2008},
  publisher={Cambridge University Press Cambridge, UK}
}

@Inbook{ramos2018fission,
author={Ramos-Fernandez, Gabriel and Aureli, Filippo},
editor={Vonk, Jennifer and Shackelford, Todd},
title={Fission-Fusion},
bookTitle={Encyclopedia of Animal Cognition and Behavior},
year={2018},
publisher={Springer International Publishing},
address={Cham},
pages={1--8},
doi={10.1007/978-3-319-47829-6_1881-1}}

@article{sole2019liquid,
author = {Solé, Ricard  and Moses, Melanie  and Forrest, Stephanie },
title = {Liquid brains, solid brains},
journal = {Philosophical Transactions of the Royal Society B: Biological Sciences},
volume = {374},
number = {1774},
pages = {20190040},
year = {2019},
doi = {10.1098/rstb.2019.0040}
}

@article{vining2019mobility,
author = {Vining, William F.  and Esponda, Fernando  and Moses, Melanie E.  and Forrest, Stephanie },
title = {How does mobility help distributed systems compute?},
journal = {Philosophical Transactions of the Royal Society B: Biological Sciences},
volume = {374},
number = {1774},
pages = {20180375},
year = {2019},
doi = {10.1098/rstb.2018.0375}
}

@article{galef2001social,
title = {Social influences on foraging in vertebrates: causal mechanisms and adaptive functions},
journal = {Animal Behaviour},
volume = {61},
number = {1},
pages = {3-15},
year = {2001},
issn = {0003-3472},
doi = {https://doi.org/10.1006/anbe.2000.1557},
author = {Bennett G. Galef and Luc-Alain Giraldeau}
}

@book{baddeley2015spatial,
  title={Spatial point patterns: methodology and applications with R},
  author={Baddeley, Adrian and Rubak, Ege and Turner, Rolf},
  year={2015},
  publisher={CRC press}
}

@article{hills2015exploration,
title = {Exploration versus exploitation in space, mind, and society},
journal = {Trends in Cognitive Sciences},
volume = {19},
number = {1},
pages = {46-54},
year = {2015},
doi = {https://doi.org/10.1016/j.tics.2014.10.004},
author = {Thomas T. Hills and Peter M. Todd and David Lazer and A. David Redish and Iain D. Couzin}
}

@article{luppi2024information,
  title={Information decomposition and the informational architecture of the brain},
  author={Luppi, Andrea I and Rosas, Fernando E and Mediano, Pedro AM and Menon, David K and Stamatakis, Emmanuel A},
  journal={Trends in Cognitive Sciences},
  year={2024},
  publisher={Elsevier}
}

@article{sonerud2001ignorant,
  title={Ignorant hooded crows follow knowledgeable roost-mates to food: support for the information centre hypothesis},
  author={Sonerud, Geir A and Smedshaug, Christian A and Br{\aa}then, {\O}ystein},
  journal={Proceedings of the Royal Society of London. Series B: Biological Sciences},
  volume={268},
  number={1469},
  pages={827--831},
  year={2001},
  publisher={The Royal Society}
}

@article{valone2002public,
  title={Public information for the assessment of quality: a widespread social phenomenon},
  author={Valone, Thomas J and Templeton, Jennifer J},
  journal={Philosophical Transactions of the Royal Society of London. Series B: Biological Sciences},
  volume={357},
  number={1427},
  pages={1549--1557},
  year={2002},
  publisher={The Royal Society}
}

@article{clark1984foraging,
  title={Foraging and flocking strategies: information in an uncertain environment},
  author={Clark, Colin W and Mangel, Marc},
  journal={The American Naturalist},
  volume={123},
  number={5},
  pages={626--641},
  year={1984},
  publisher={University of Chicago Press}
}

@article{petit2010decision,
  title={Decision-making processes: the case of collective movements},
  author={Petit, Odile and Bon, Richard},
  journal={Behavioural processes},
  volume={84},
  number={3},
  pages={635--647},
  year={2010},
  publisher={Elsevier}
}

@article{bettencourt2009rules,
  title={The rules of information aggregation and emergence of collective intelligent behavior},
  author={Bettencourt, Lu{\'\i}s MA},
  journal={Topics in Cognitive Science},
  volume={1},
  number={4},
  pages={598--620},
  year={2009},
  publisher={Wiley Online Library}
}

@book{gudhi2015urm
, title        = "{GUDHI} User and Reference Manual"
, author      = "{The GUDHI Project}"
, publisher     = "{GUDHI Editorial Board}"
, year         = 2015
}

@article{lyons2013home,
  title={Home range plus: a space-time characterization of movement over real landscapes},
  author={Lyons, Andrew J and Turner, Wendy C and Getz, Wayne M},
  journal={Movement Ecology},
  volume={1},
  pages={1--14},
  year={2013},
  publisher={Springer}
}

@misc{Serrano2020,
      title={Centrality measures in simplicial complexes: applications of Topological Data Analysis to Network Science}, 
      author={Hernández Serrano, D and Sánchez Gómez, D},
      year={2020},
      eprint={1908.02967},
      archivePrefix={arXiv},
      primaryClass={math.AT},
      note={Preprint}
}

@misc{garg2025causal,
    title  = {The causal role of synergy in collective problem-solving},
    author={Garg, Ketika and Moser, Cody and Dromiack, Hannah and Anwarzai, Zara and Ramos-Fernandez, Gabriel},
    year={2025},
    eprint={https://osf.io/6r2h5_v2},
    archivePrefix={SocArXiv},
    primaryClass={},
    note={Preprint}
}

@article{hasenjager2024social,
  title={Social ageing and higher-order interactions: social selectiveness can enhance older individuals’ capacity to transmit knowledge},
  author={Hasenjager, Matthew J and Fefferman, Nina H},
  journal={Philosophical Transactions B},
  volume={379},
  number={1916},
  pages={20220461},
  year={2024},
  publisher={The Royal Society}
}

@article{silk2023conceptual,
  title={Conceptual representations of animal social networks: an overview},
  author={Silk, Matthew J},
  journal={Animal Behaviour},
  volume={201},
  pages={157--166},
  year={2023},
  publisher={Elsevier}
}

@article{Iacopini2024Temporal,
  title = {The temporal dynamics of group interactions in higher-order social networks},
  volume = {15},
  DOI = {10.1038/s41467-024-50918-5},
  number = {1},
  journal = {Nature Communications},
  publisher = {Springer Science and Business Media LLC},
  author = {Iacopini,  Iacopo and Karsai,  Márton and Barrat,  Alain},
  year = {2024}
}

@article{Lin2024,
  title = {Higher-order non-Markovian social contagions in simplicial complexes},
  volume = {7},
  DOI = {10.1038/s42005-024-01666-x},
  number = {1},
  journal = {Communications Physics},
  publisher = {Springer Science and Business Media LLC},
  author = {Lin,  Zhaohua and Han,  Lilei and Feng,  Mi and Liu,  Ying and Tang,  Ming},
  year = {2024}
}

@article{Iacopini2024,
  title = {Not your private t\^ete-à-t\^ete: leveraging the power of higher-order networks to study animal communication},
  volume = {379},
  DOI = {10.1098/rstb.2023.0190},
  number = {1905},
  journal = {Philosophical Transactions of the Royal Society B: Biological Sciences},
  publisher = {The Royal Society},
  author = {Iacopini,  Iacopo and Foote,  Jennifer R. and Fefferman,  Nina H. and Derryberry,  Elizabeth P. and Silk,  Matthew J.},
  year = {2024}
}

@article{Otter2017,
  title = {A roadmap for the computation of persistent homology},
  volume = {6},
  ISSN = {2193-1127},
  DOI = {10.1140/epjds/s13688-017-0109-5},
  number = {1},
  journal = {EPJ Data Science},
  publisher = {Springer Science and Business Media LLC},
  author = {Otter,  Nina and Porter,  Mason A. and Tillmann,  Ulrike and Grindrod,  Peter and Harrington,  Heather A.},
  year = {2017},
  month = aug 
}

@article{Silk2022,
author = {Silk, Matthew J. and Wilber, Mark Q. and Fefferman, Nina H.},
title = {Capturing complex interactions in disease ecology with simplicial sets},
journal = {Ecology Letters},
volume = {25},
number = {10},
pages = {2217-2231},
doi = {https://doi.org/10.1111/ele.14079},
year = {2022}
}

@article{getz2007locoh,
    doi = {10.1371/journal.pone.0000207},
    author = {Getz, Wayne M. AND Fortmann-Roe, Scott AND Cross, Paul C. AND Lyons, Andrew J. AND Ryan, Sadie J. AND Wilmers, Christopher C.},
    journal = {PLOS ONE},
    publisher = {Public Library of Science},
    title = {LoCoH: Nonparameteric Kernel Methods for Constructing Home Ranges and Utilization Distributions},
    year = {2007},
    month = {02},
    volume = {2},
    pages = {1-11},
    number = {2},
}

@Manual{rcoreteam2021R,
     title = {R: A Language and Environment for Statistical Computing},
    author = {{R Core Team}},
    organization = {R Foundation for Statistical Computing},
    address = {Vienna, Austria},
    year = {2021}
  }

@mastersthesis{pinacho2010patrones,
    author = {Pinacho Guendulain, Braulio},
    title = {Patrones de agrupación de un grupo de monos araña de manos negras (Ateles geoffroyi) en Punta Laguna, Yucatán},
    school = {Instituto Politécnico Nacional},
    year = {2010},
http = {http://literatura.ciidiroaxaca.ipn.mx:8080/xmlui/handle/LITER_CIIDIROAX/112}
}

@article{janmaat2016spatio,
  title={Spatio-temporal complexity of chimpanzee food: How cognitive adaptations can counteract the ephemeral nature of ripe fruit},
  author={Janmaat, Karline RL and Boesch, Christophe and Byrne, Richard and Chapman, Colin A and Gon{\'e} Bi, Zoro B and Head, Josephine S and Robbins, Martha M and Wrangham, Richard W and Polansky, Leo},
  journal={American Journal of Primatology},
  volume={78},
  number={6},
  pages={626--645},
  year={2016},
  publisher={Wiley Online Library}
}

@phdthesis{ramos2001patterns,
    author = {Ramos-Fernández, Gabriel},
    title = {Patterns of association, feeding competition and vocal communication in spider monkeys, Ateles geoffroyi},
    school = {University of Pennsylvania},
    year = {2001}
}

@phdthesis{bonilla2008forest,
  title={Forest recovery and management options in the Yucatan Peninsula, Mexico},
  author={Bonilla Moheno, Martha},
  year={2008},
  school={University of California, Santa Cruz}
}

@article{smith2016seasonal,
  title={Seasonal changes in socio-spatial structure in a group of free-living spider monkeys (Ateles geoffroyi)},
  author={Smith-Aguilar, Sandra E and Ramos-Fern{\'a}ndez, Gabriel and Getz, Wayne M},
  journal={PloS one},
  volume={11},
  number={6},
  pages={e0157228},
  year={2016},
  publisher={Public Library of Science San Francisco, CA USA}
}

@book{hutchins1995cognition,
  title={Cognition in the Wild},
  author={Hutchins, Edwin},
  year={1995},
  publisher={MIT press}
}

@article{williams2010nonnegative,
  title={Nonnegative decomposition of multivariate information},
  author={Williams, Paul L and Beer, Randall D},
  journal={arXiv preprint arXiv:1004.2515},
  year={2010}
}

@article{ramos2020collective,
AUTHOR={Ramos-Fernandez, Gabriel and Smith Aguilar, Sandra E. and Krakauer, David C. and Flack, Jessica C.},   
TITLE={Collective Computation in Animal Fission-Fusion Dynamics},     JOURNAL={Frontiers in Robotics and AI},      
VOLUME={7},           
YEAR={2020},      	  
DOI={10.3389/frobt.2020.00090},      	
ISSN={2296-9144},   
}

@article{aureli2008fission,
  title={Fission-fusion dynamics: new research frameworks},
  author={Aureli, Filippo and Schaffner, Colleen M and Boesch, Christophe and Bearder, Simon K and Call, Josep and Chapman, Colin A and Connor, Richard and Fiore, Anthony Di and Dunbar, Robin IM and Henzi, S Peter and others},
  journal={Current Anthropology},
  volume={49},
  number={4},
  pages={627--654},
  year={2008},
  publisher={The University of Chicago Press}
}

@ARTICLE{SciPy2020,
  author  = {Virtanen, Pauli and Gommers, Ralf and Oliphant, Travis E. and
            Haberland, Matt and Reddy, Tyler and Cournapeau, David and
            Burovski, Evgeni and Peterson, Pearu and Weckesser, Warren and
            Bright, Jonathan and {van der Walt}, St{\'e}fan J. and
            Brett, Matthew and Wilson, Joshua and Millman, K. Jarrod and
            Mayorov, Nikolay and Nelson, Andrew R. J. and Jones, Eric and
            Kern, Robert and Larson, Eric and Carey, C J and
            Polat, {\.I}lhan and Feng, Yu and Moore, Eric W. and
            {VanderPlas}, Jake and Laxalde, Denis and Perktold, Josef and
            Cimrman, Robert and Henriksen, Ian and Quintero, E. A. and
            Harris, Charles R. and Archibald, Anne M. and
            Ribeiro, Ant{\^o}nio H. and Pedregosa, Fabian and
            {van Mulbregt}, Paul and {SciPy 1.0 Contributors}},
  title   = {{{SciPy} 1.0: Fundamental Algorithms for Scientific
            Computing in Python}},
  journal = {Nature Methods},
  year    = {2020},
  volume  = {17},
  pages   = {261--272},
  adsurl  = {https://rdcu.be/b08Wh},
  doi     = {10.1038/s41592-019-0686-2},
}

@article{suarez2006diet,
  title={Diet and travel costs for spider monkeys in a nonseasonal, hyperdiverse environment},
  author={Suarez, Scott A},
  journal={International Journal of Primatology},
  volume={27},
  pages={411--436},
  year={2006},
  publisher={Springer}
}

@article{marshall2024aims,
  title={On aims and methods of collective animal behaviour},
  author={Marshall, James AR and Reina, Andreagiovanni},
  journal={Animal Behaviour},
  volume={210},
  pages={189--197},
  year={2024},
  publisher={Elsevier}
}

@incollection{wallace2008factors,
  title={Factors influencing spider monkey habitat use and ranging patterns},
  author={Wallace, Robert B},
   booktitle={Spider monkeys: Behavior, ecology and evolution of the genus Ateles},
  editor = {Campbell, Christina J},
  pages={138--154},
  year={2008},
  publisher={Cambridge University Press Cambridge, UK}
}

@article{ramos2013site,
    doi = {10.1371/journal.pone.0062813},
    author = {Ramos-Fernandez, Gabriel AND Smith Aguilar, Sandra E. AND Schaffner, Colleen M. AND Vick, Laura G. AND Aureli, Filippo},
    journal = {PLOS ONE},
    publisher = {Public Library of Science},
    title = {Site Fidelity in Space Use by Spider Monkeys (Ateles geoffroyi) in the Yucatan Peninsula, Mexico},
    year = {2013},
    month = {05},
    volume = {8},
    pages = {1-10},
    number = {5}
}

@article{moussaid2009collective,
  title={Collective information processing and pattern formation in swarms, flocks, and crowds},
  author={Moussaid, Mehdi and Garnier, Simon and Theraulaz, Guy and Helbing, Dirk},
  journal={Topics in Cognitive Science},
  volume={1},
  number={3},
  pages={469--497},
  year={2009},
  publisher={Wiley Online Library}
}

@article{palacios2019uncovering,
  title={Uncovering the decision rules behind collective foraging in spider monkeys},
  author={Palacios-Romo, TM and Castellanos, F and Ramos-Fernandez, G},
  journal={Animal Behaviour},
  volume={149},
  pages={121--133},
  year={2019},
  publisher={Elsevier}
}

@article{pinacho2017influence,
  title={Influence of fruit availability on the fission--fusion dynamics of spider monkeys (\emph{{A}teles geoffroyi})},
  author={Pinacho-Guendulain, Braulio and Ramos-Fern{\'a}ndez, Gabriel},
  journal={International Journal of Primatology},
  volume={38},
  number={3},
  pages={466--484},
  year={2017},
  publisher={Springer}
}

@article{ramos2009association,
  title={Association networks in spider monkeys (Ateles geoffroyi)},
  author={Ramos-Fern{\'a}ndez, Gabriel and Boyer, Denis and Aureli, Filippo and Vick, Laura G},
  journal={Behavioral Ecology and Sociobiology},
  volume={63},
  number={7},
  pages={999--1013},
  year={2009},
  publisher={Springer}
}

@article{ramos2014unraveling,
  title={Unraveling fission-fusion dynamics: how subgroup properties and dyadic interactions influence individual decisions},
  author={Ramos-Fern{\'a}ndez, Gabriel and Morales, Juan M},
  journal={Behavioral Ecology and Sociobiology},
  volume={68},
  number={8},
  pages={1225--1235},
  year={2014},
  publisher={Springer}
}

@article{ramos2018quantifying,
  title={Quantifying uncertainty due to fission--fusion dynamics as a component of social complexity},
  author={Ramos-Fern\'andez, Gabriel and King, Andrew J and Beehner, Jacinta C and Bergman, Thore J and Crofoot, Margaret C and Di Fiore, Anthony and Lehmann, Julia and Schaffner, Colleen M and Snyder-Mackler, Noah and Zuberb{\"u}hler, Klaus and others},
  journal={Proceedings of the Royal Society B: Biological Sciences},
  volume={285},
  number={1879},
  pages={20180532},
  year={2018},
  publisher={The Royal Society}
}

@article{ramos2018ecologia,
  title={Ecolog{\'\i}a, comportamiento y conservaci{\'o}n de los monos ara{\~n}a (Ateles geoffroyi): 20 a{\~n}os de estudio en Punta Laguna, M{\'e}xico},
  author={Ramos-Fern{\'a}ndez, Gabriel and Aureli, Filippo and Schaffner, Colleen M and Vick, Laura G},
  journal={La primatolog{\'\i}a en Latinoam{\'e}rica},
  volume={2},
  pages={531--544},
  year={2018}
}

@book{kummer2006primate,
  title={Primate societies: Group techniques of ecological adaptation},
  author={Kummer, Hans},
  year={2006},
  publisher={Transaction Publishers}
}

@article{king2011next,
  title={Where next? Group coordination and collective decision making by primates},
  author={King, Andrew J and Sueur, C{\'e}dric},
  journal={International journal of primatology},
  volume={32},
  pages={1245--1267},
  year={2011},
  publisher={Springer}
}

@article{sueur2011collective,
  title={Collective decision-making and fission--fusion dynamics: a conceptual framework},
  author={Sueur, C{\'e}dric and King, Andrew J and Conradt, Larissa and Kerth, Gerald and Lusseau, David and Mettke-Hofmann, Claudia and Schaffner, Colleen M and Williams, Leah and Zinner, Dietmar and Aureli, Filippo},
  journal={Oikos},
  volume={120},
  number={11},
  pages={1608--1617},
  year={2011},
  publisher={Wiley Online Library}
}

@article{falcon2019collective,
  title={Collective learning from individual experiences and information transfer during group foraging},
  author={Falc{\'o}n-Cort{\'e}s, Andrea and Boyer, Denis and Ramos-Fern{\'a}ndez, Gabriel},
  journal={Journal of the Royal Society Interface},
  volume={16},
  number={151},
  pages={20180803},
  year={2019},
  publisher={The Royal Society}
}
\section*{Acknowledgements}
This work was partly conducted during a sabbatical stay by GRF at the Global Research Centre for Diverse Intelligences at the University of St. Andrews, supported by a PASPA-DGAPA grant from the Universidad Nacional Autónoma de México. RSW was supported by the EPSRC Centre for Doctoral Training in Mathematical Modelling, Analysis and Computation (MAC-MIGS) funded by the UK Engineering and Physical Sciences Research Council (grant EP/S023291/1). SSA was supported by a postdoctoral grant (\textit{Estancias posdoctorales por México}) from the Secretaría de Ciencia, Humanidades, Tecnología e Innovación (SECIHTI). Fieldwork was carried out under research permits DGVS00910/13 and DGVS02716/14 from the Secretaría de Medio Ambiente y Recursos Naturales (SEMARNAT). We thank Lev Guzman-Vargas, Alvaro Zabaleta-Ortega and two anonymous reviewers for their helpful comments on previous versions of the manuscript.

\section*{Author Information}
\subsection*{Authors and affiliations}
\subsubsection*{Instituto de Investigaciones en Matemáticas Aplicadas y en Sistemas, Universidad Nacional Autónoma de México}
Gabriel Ramos-Fernandez \\
Sandra E. Smith Aguilar
\subsubsection*{Global Research Centre for Diverse Intelligences, University of St. Andrews}
Gabriel Ramos-Fernandez
\subsubsection*{Department of Mathematics and Maxwell Institute for Mathematical Sciences, Heriot-Watt University}
Ross S. Walker
\subsubsection*{Institute of Ecology and Evolution, University of Edinburgh} 
Matthew J. Silk \\
Ross S. Walker
\subsubsection*{Instituto de Física, Universidad Nacional Autónoma de México}
Denis Boyer
\subsection*{Contributions}
Gabriel Ramos-Fernandez: Conceptualization, Methodology, Formal analysis, Data curation, Writing - original draft, Writing - review \& editing, Visualization, Supervision, Project administration, Funding acquisition. Ross S. Walker: Conceptualization, Methodology, Formal analysis, Software, Validation, Writing - original draft, Writing - review \& editing, Visualization. Matthew J. Silk:  Conceptualization, Methodology, Formal analysis, Writing - review \& editing, Supervision. Denis Boyer: Conceptualization, Methodology, Formal analysis, Writing - review \& editing. Sandra E. Smith Aguilar: Conceptualization, Methodology, Formal analysis, Data curation, Writing - original draft, Writing - review \& editing, Visualization. 
\subsection*{Corresponding authors}
Correspondence to Gabriel Ramos-Fernandez.
\section*{Ethics declarations}
\subsection*{Competing interests}
All authors declare no financial or non-financial competing interests.

\title{Supplementary information for
\\
Uncovering complementary information sharing in spider monkey collective foraging using higher-order spatial networks}
\author{Gabriel Ramos-Fernandez, Ross S. Walker, Matthew J. Silk, Denis Boyer and Sandra E. Smith Aguilar}

%\date{\today}

\maketitle

This document contains further methodological detail on the study subjects and the derivation of the optimal relative spatial overlap in section 1, while section 2 contains additional results on: 1) the correlation between the filtration complementarity index and the seasons and the variation in fruit abundance; and 2) the maximal simplicial centrality analysis and the individual composition of simplices.

\section{Methods}
\subsection{Further data details}
\begin{figure}[H]
    \centering
    \includegraphics[trim={0 0 0 0}, clip, width=1.0\textwidth]{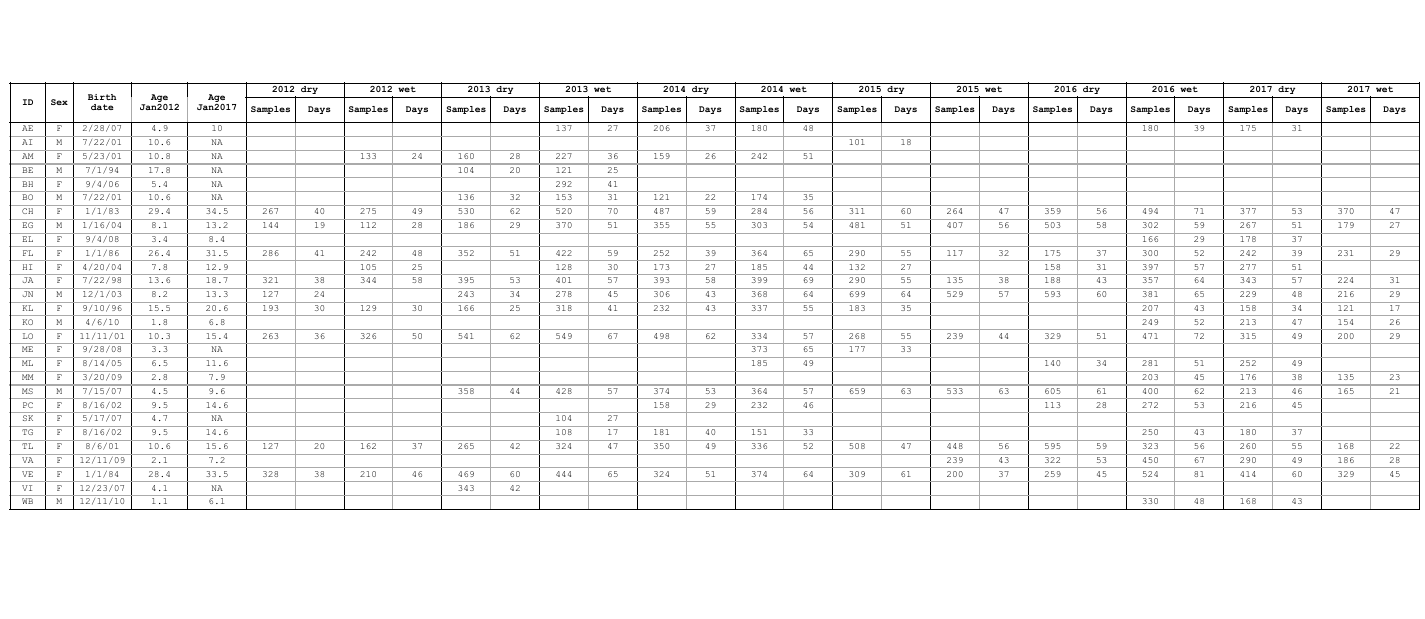}\\[1mm]
  {Table SI1: sample size, in terms of the number of scan samples and days of observation, in each season for each individual spider monkey included in the study. Also shown are the sex and date of birth, as well as the age of each individual in the first and last years of this study.}
\label{fig:tableSI}
\end{figure}

\subsection{Optimal space sharing and subgroup size}
In this subsection we pose and prove the result that the optimal relative spatial overlap between all individuals (in a subgroup of size $n$) is exactly 
\begin{equation}\label{eq: main_optim_result}
  w_n^* = \frac{1}{n+1}
\end{equation}
In the main paper this result was used to determine how redundant were the spatial interactions between core ranges observed from the data when taking into account subgroup size. This section first constructs a suitable objective function which describes the total level of information transfer happening in a subgroup of $n$ individuals under a particular spatial overlap structure. Then, assuming that each individual has equal foraging ability, we prove global optimality of a particular spatial structure which corresponds to the result \eqref{eq: main_optim_result}. 

\subsubsection{Construction of an objective function}
We want to find the `area overlap' between home ranges which maximises the information transfer between $n$ individuals. For this, we need to construct an objective function (the \textit{objective function}) and a suitable set of constraints on the optimisation problem. 

The union of the core ranges of all individuals are partitioned into areas known uniquely by distinct subsets of individuals. The areas which are known by multiple individuals will be the variables of the objective function.  Throughout, we make the following assumptions: 
\begin{enumerate}
    \item Individuals utilise the area in their core ranges uniformly.
    \item core ranges are composed of cells, representing, for example, points of knowledge. 
    \item Movements of different individuals are independent.
    \item Rate of information transfer for a given grouping of individuals is given by the number of cells known by the group known by at least one member of the group and not known by at least one member, multiplied by the probability of the interaction occurring.  
\end{enumerate}

Suppose we have $n$-many individuals, enumerated from $1$ to $n$. By Assumption 2, the $i$-th individual has a core range consisting of $N_i$ many cells for each $i=1,\dots, n$. Denote by $[n]$ the set $\{1,\dots, n\}$. Denote the power set of $[n]$ by $\mathcal{P}([n])$. Then, the set of all multi-individual interactions, denoted $\mathcal{X}$, will be $\mathcal{P}([n])$ excluding singletons and the empty set $\emptyset$. The size of this set, $\mathcal{N}=|\mathcal{X}|$, will then be 
\begin{equation*}
    \mathcal{N} = \underbrace{2^n}_{\text{size of }\mathcal{P}([n])}-\underbrace{n}_{\text{number of singletons in }\mathcal{P}([n])}-\underbrace{1}_{\text{the empty set }\emptyset}
\end{equation*}
(which implies that the optimisation problem will be $2^n-n-1$ dimensional). Each element of $\chi$ will be of the form $\{i_1, ..., i_k\}$ where each $i_l\in [n]$. We can arrange the entries of each of these elements so $i_1, ..., i_k$ is an increasing sequence. Then we can then place an ordering upon $\mathcal{X}$ by arranging the elements in ascending cardinality (so interactions between fewer individuals come before interactions between more individuals) and then order elements of the \textit{same} cardinality lexicographically. Some examples: $\{1,2\}$ will be the first element in the ordering, $\{1, 2, 3\}$ will be the first element of size 3 and $[n]$ (the set of all individuals) will be the final element. This ordering allows us to enumerate the overlapping areas according to their position in the ordering of $\mathcal{X}$. We represent the enumeration with the (bijective) mapping $I\colon \mathcal{X} \to \left[\mathcal{N}\right]$, which matches a group of individuals with its index in the ordering. Equivalently, for a given $c\in \left [\mathcal{N} \right ]$, $I^{-1}(c)$ gives the $c$-th group of individuals in the ordering of $\mathcal{X}$. With this construction we can list all of our area overlaps (the variables of our objective function) as $\left(O_1, \dots, O_\mathcal{N}\right)$. Note that this particular choice of ordering was chosen to maintain interpretability of results and ensure easy compatibility with optimisation software, although the choice is mathematically arbitrary.

We account for the fact that any grouping of individuals can interact in both the areas that they share uniquely \textit{and} in the areas shared by them and additional individuals. For example, individuals 1 and 2 can meet in the area shared by individuals 1, 2 and 3 without the presence of individual 3. This motivates the following definition. Let $S\colon\left[\mathcal{N}\right]\to\mathcal{P}\left( \left[\mathcal{N}\right]\right)$ be a mapping such that $S(c)$ is the collection of indices of areas known by the individuals $I^{-1}(c)$. We can additionally write 
\begin{equation*}
    S(c) = \left\{i\in \left[\mathcal{N} \right] : I^{-1}(c)\subseteq I^{-1}(i)\right\}.
\end{equation*}
For example, in the case $n=3$, we have $\mathcal{X}=\{\{1, 2\}, \{1, 3\}, \{2, 3\}, \{1,2,3\}\}$, with ordering map given by $I(\{1, 2\}) = 1$, $I(\{1, 3\}) = 2$, $I(\{2, 3\}) = 3$ and $I(\{1, 2, 3\}) = 4$. The corresponding $S$ mapping is $S(1) = \{1, 4\}$, $S(2) = \{2, 4\}$, $S(3) = \{3, 4\}$, $S(4) = \{4\}$. With these constructions we can finally write that, for a given group of individuals, with index $c\in \left[\mathcal{N}\right]$, the probability of them meeting is given by 
\begin{equation*}
    P\left(c, O_1, \dots, O_{\mathcal{N}}\right) = \underbrace{\left(\prod_{k\in I^{-1}(c)}N_k\right)^{-1}}_{\text{Probability of meeting in same cell}} \cdot \underbrace{\left( \sum_{a\in S(c)} O_a\right)}_{\text{Number of cells for meeting}}
\end{equation*}
when applying Assumptions 1 (uniform space use) and 3 (independence of movements). 

Now we quantify the amount of unique information known by each subgroup in a way that aligns with Assumption 4. We consider the total number of cells in each of the core ranges, and then remove the number of cells known by all individuals, and ensure that all cells which is not known by all individuals are not over counted. To do this efficiently, we need to introduce more notation. 

Let $f\colon \left[\mathcal{N}\right]\to \mathbb{R}$ be given by $f(c) = |I^{-1}(c)|$ by the number of individuals in the $c$-th grouping corresponding to the ordering given by $I$. Additionally, for brevity of notation, let $g\colon \left[\mathcal{N}\right]\to \mathbb{R}$ be given by $g(c)=f(c)-1$. Define $B\colon \left[\mathcal{N} \right]\to \mathcal{P}\left(\left[\mathcal{N} \right]\right)$ such that $B(c)$ is the set of indices of all the smaller collections of individuals contained in the $c$-th collection of individuals:
\begin{equation*}
B(c) = \left\{i\in \left[\mathcal{N} \right] : I^{-1}(i) \subset I^{-1}(c)\right\},
\end{equation*}
where $\subset$ is understood to mean \textit{strict} containment. Then the amount of unique area known by the $c$-th subset of individuals can be written as 
\begin{equation*}
    A\left(c, O_1, \dots, O_{\mathcal{N}}\right) = \underbrace{\sum_{k\in I^{-1}(c)}N_k}_{\text{Sum of all areas}} - \underbrace{f(c)\sum_{a\in S(c)}O_a}_{\text{Area shared by all individuals}}-\underbrace{\sum_{a\in B(c)}g(a)O_a}_{\text{Areas shared by subgroups}}
\end{equation*}
where the second sum removes the number of cells which are known by all individuals in the group which are used for the transfer of information (which will be included $f(c)$ many times in the first sum) and the final sum ensures there is no over-counting of uniquely known cells (by removing all but one of the occasions upon which those cells were added to $A$). Then by Assumption 4, we may write the objective function as 
\begin{align*}
    T\left(O_1,\dots, O_\mathcal{N}\right) &= \sum_{c=1}^{\mathcal{N}}P\left(c, O_1,\dots, O_\mathcal{N}\right)A\left(c, O_1,\dots, O_\mathcal{N}\right) \\
    &= \sum_{c=1}^{\mathcal{N}}\left[\left( \frac{\sum_{a\in S(c)} O_a}{\prod_{k\in I^{-1}(c)}N_k}\right) \left(\sum_{k\in I^{-1}(c)}N_k -f(c)\sum_{a\in S(c)}O_a-\sum_{a\in B(c)}g(a)O_a \right)\right]
    %&=\sum_{c=1}^{\mathcal{N}}\left[\frac{\sum_{k\in I^{-1}(c)}N_k}{\prod_{k\in I^{-1}(c)}N_k}\sum_{a\in S(c)}O_a-\frac{1}{{\prod_{k\in I^{-1}(c)}N_k}}\left(f(c)O_c\sum_{a\in S(c)}O_a +\sum_{(a_1 ,a_2)\in B(c)\times S(c)} g(O_{a_1})O_{a_1}O_{a_2}\right)\right]
\end{align*}
which describes the net amount of information transfer. This is the sum of products of a linear function with an (affine) linear function, which implies that $T$ is a quadratic form in $\mathcal{N}$ variables. Therefore, we can write $T$ \textit{uniquely} in the form 
\begin{equation}\label{eq:objective}
    T\left(\boldsymbol{O}\right) = \boldsymbol{l}^t\boldsymbol{O}+\frac{1}{2}\boldsymbol{O}^tM\boldsymbol{O} 
\end{equation}
where $\boldsymbol{O}\in \mathbb{R}^\mathcal{N}$ is the vector of all the $O_i$'s, $\boldsymbol{l}\in\mathbb{R}^\mathcal{N}$ and $M\in\mathbb{R}^{\mathcal{N}\times\mathcal{N}}$. We collect the terms in $T$ to determine the vector $\boldsymbol{l}$ and matrix $M$. To do this, we first break the objective function into two parts, $L$ and $Q$, which contain the linear and quadratic terms of $T$ respectively, such that
\begin{align*}
    &T\left(\boldsymbol{O}\right)=L\left(\boldsymbol{O}\right) - Q\left(\boldsymbol{O}\right), \\
    & L\left(\boldsymbol{O}\right) = \sum_{c=1}^{\mathcal{N}}\left[\frac{\sum_{k\in I^{-1}(c)}N_k}{\prod_{k\in I^{-1}(c)}N_k}\sum_{a\in S(c)}O_a\right], \\ 
    & Q\left(\boldsymbol{O}\right)=\sum_{c=1}^{\mathcal{N}}\left[\frac{1}{\prod_{k\in I^{-1}(c)}N_k} \left(f(c)\sum_{a_1, a_2\in S(c)}O_{a_1}O_{a_2}+ \sum_{\substack{a_1 \in B(c)\\ a_2 \in S(c)}} g\left(a_1\right)O_{a_1}O_{a_2}\right)\right].
\end{align*}

We can determine the entries of $\boldsymbol{l}$ by finding the coefficients of the linear terms $O_i$ in $T\left(\boldsymbol{O}\right)$ for each $i\in \left[\mathcal{N}\right]$, which will be contained in $L\left(\boldsymbol{O}\right)$. Let $i\in \left[\mathcal{N}\right]$. Then there is a contribution from the first sum (over $c$) to the coefficient of $O_i$ only for $c\in \left[\mathcal{N}\right]$ such that $i\in S(c)$. For conciseness, we define $J\colon \left[\mathcal{N}\right]\to \mathcal{P}(\left[\mathcal{N}\right])$ as 
\begin{equation*}
    J(i) =\left\{c\in \left[\mathcal{N}\right] : i\in S(c)\right\}.
\end{equation*}
This allows us to write the coefficient of $O_i$ in $L(\boldsymbol{O})$ as 
\begin{equation*}
    \operatorname{Coeff}(O_i) = \sum_{c\in J(i)} \left[\frac{\sum_{k\in I^{-1}(c)}N_k}{\prod_{k\in I^{-1}(c)}N_k}\right]
\end{equation*}
such that we can determine the entries of $\boldsymbol{l} = (l_i)$ as $l_i = \operatorname{Coeff}(O_i)$.

We can also find the entries of the matrix $Q$ by examining the coefficients of terms of the form $O_iO_j$ in $T\left(\boldsymbol{O}\right)$ for all $i, j \in \left[\mathcal{N}\right]$, which will be contained in the $Q\left(\boldsymbol{O}\right)$ part of $T\left(\boldsymbol{O}\right)$. To simplify our argument, we break $Q\left(\boldsymbol{O}\right)$ into two parts
\begin{align*}
    &Q\left(\boldsymbol{O}\right) = Q_1\left(\boldsymbol{O}\right)+Q_2\left(\boldsymbol{O}\right), \\
    & Q_1\left(\boldsymbol{O}\right) = \sum_{c=1}^{\mathcal{N}}\left[\frac{1}{\prod_{k\in I^{-1}(c)}N_k} \left(f(c)\sum_{a_1, a_2\in S(c)}O_{a_1}O_{a_2}\right)\right],\\
    & Q_2\left(\boldsymbol{O}\right) = \sum_{c=1}^{\mathcal{N}}\left[\frac{1}{\prod_{k\in I^{-1}(c)}N_k} \left(\sum_{\substack{a_1 \in B(c)\\ a_2 \in S(c)}} g\left(a_1\right)O_{a_1}O_{a_2}\right) \right].
\end{align*}
Let $i, j\in\left[\mathcal{N}\right]$. We first consider contributions to coefficients from $Q_1$. Here there is a contribution to $O_iO_j$ only for the values of $c$ with $i\in S(c)$ and $j\in S(c)$, meaning the $c$ values such that $I^{-1}(c)\subseteq I^{-1}(i)$ and $I^{-1}(c)\subseteq I^{-1}(j)$. These two conditions are equivalent to the single condition $I^{-1}(c)\subseteq I^{-1}(i)\cap I^{-1}(j)$. This now motivates the definition of a new function $V\colon \left[\mathcal{N} \right]\times \left[\mathcal{N} \right] \to \left[\mathcal{N} \right]$ with 
\begin{equation*}
    V(i,j) = \left\{c: I^{-1}(c)\subseteq I^{-1}(i)\cap I^{-1}(j) \right\}.
\end{equation*}
For each $c$ in this set, we will get a contribution of $2f(c)$ to the coefficient of $O_iO_j$ in $Q_1$ if $i\neq j$ (the scaling of $2$ comes from the fact that the \textit{unordered} pair $(i,j)$ appears twice in the sum, from the cases $a_1=i$, $a_2=j$ and $a_1=j$, $a_2=i$) and a contribution of $f(c)$ if $i=j$.

This allows us to neatly write the the coefficient of $O_iO_j$ in $T$ from $Q_1$, denoted $\operatorname{Coeff}_1(O_iO_j)$, as
\begin{equation*}
    \operatorname{Coeff}_1(O_iO_j) = \begin{cases}
         \sum_{c\in V(i,j)}\frac{2f(c)}{\prod_{k\in I^{-1}(c)}N_k}\text{, if } i\neq j \\ \sum_{c\in V(i,j)}\frac{f(c)}{\prod_{k\in I^{-1}(c)}N_k}\text{, if } i= j
    \end{cases}
\end{equation*}

Now we consider the contributions from $Q_2$. The $c$ values for which the term $O_iO_j$ appears are those such that $i\in S(c)$ and $j\in B(c)$ (or the other way around, which is handled analogously). This is equivalent to the relevant $c$ values being those such that $I^{-1}(c)\subseteq I^{-1}(i)$ and $I^{-1}(j)\subset I^{-1}(c)$. Note that this condition implies, by transitivity, that $I^{-1}(j) \subset I^{-1}(i)$, such that if this condition (or the reverse) is not satisfied, the contribution from this sum will be zero. That this implies all terms $O_iO_j$ with $i=j$ will not appear in $Q_2$. To collect all $c$ values satisfying these two conditions, we define $Z\colon \left[\mathcal{N}\right]\times \left[\mathcal{N}\right]\to \left[\mathcal{N}\right]$ such that 
\begin{equation*}
    Z(i: j) = \left\{c : I^{-1}(j)\subset I^{-1}(c)\subseteq I^{-1}(i) \right\}
\end{equation*}
where the colon notation is used to emphasise that $Z(i:j)\neq Z(j:i)$. %Note that the size of the set $Z(i:j)$ will be $2^{f(i)-f(j)}-1$, whenever it is non-empty. 
This construction allows us to write the contribution from $Q_2$ as 
\begin{equation*}
    \operatorname{Coeff_2}(O_iO_j) = \begin{cases}
        \sum_{c\in Z(i: j)}\frac{g(j)}{\prod_{k\in I^{-1}(c)}N_k}\text{, if } I^{-1}(j)\subset I^{-1}(i),  \\ 
        \sum_{c\in Z(j: i)}\frac{g(i)}{\prod_{k\in I^{-1}(c)}N_k}\text{, if } I^{-1}(i)\subset I^{-1}(j), \\ 0\text{, otherwise }
    \end{cases}
\end{equation*} 
where $\operatorname{Coeff_2}(O_iO_j)$ is the contribution of $Q_2$ to the coefficient of $O_iO_j$ in $T$. Finally, we add the contributions from $Q_1$ and $Q_2$ to determine the final coefficients of the second order terms in $T$ as $\operatorname{Coeff}(O_iO_j) =\operatorname{Coeff}_1(O_iO_j)+\operatorname{Coeff}_2(O_iO_j)$. This collection of coefficients allows us to `neatly' give the entries of the matrix $M = (M_{ij})$ as
\begin{equation*}
    M_{ij} = \begin{cases}
        -2\operatorname{Coeff}(O_iO_j)\text{, if } i=j \\
        -\operatorname{Coeff}(O_iO_j)\text{, otherwise }
    \end{cases}
\end{equation*}
where the sign is reversed because $Q(\boldsymbol{O})$ has a coefficient of $-1$ in $T(\boldsymbol{O})$. Note, therefore, that all entries of $M_{ij}$ will be non-positive. We note here for the next section, that since we double the diagonal terms (where $i=j$), then the effective contribution from $\operatorname{coeff}_1(O_iO_j)$ to $M$ is always 
\begin{equation*}
    m_{ij} = -\sum_{c\in V(i,j)}\frac{2f(c)}{\prod_{k\in I^{-1}(c)}N_k}
\end{equation*}

We now have the objective form written in the standard quadratic form $T\left(\boldsymbol{O}\right) = \boldsymbol{l}^t\boldsymbol{O}+\frac{1}{2}\boldsymbol{O}^tM\boldsymbol{O}$. We now need to impose some constraints on the problem to ensure that the solution makes biological sense. In particular,
\begin{itemize}
    \item All areas should be non-negative.
    \item The sum of all cells in which an individual has shared knowledge of is less than their total area.
\end{itemize}
Mathematically these conditions can be written as
\begin{itemize}
    \item $\boldsymbol{O}\geq 0$, meaning that $O_i\geq 0$, $\forall i \in \left[\mathcal{N}\right]$.
    \item $\sum_{i\in S\left(I(\{k\})\right)}O_i \leq N_k$, $\forall k \in \left[n\right]$.
\end{itemize}
The second condition (no over-sharing) is best represented in the form $G\boldsymbol{O}\leq \boldsymbol{h}$ for some matrix $G\in \mathbb{R}^{n\times \mathcal{N}}$ and some vector $h\in\mathbb{R}^n$. Thankfully, these are easier to construct than the previous matrix and vector. The vector $\boldsymbol{h}=(h_i)$ will have entries $h_i=N_i$, and the matrix $G=(G_{ij})$ will have entries 
\begin{equation*}
        G_{ij} = 
    \begin{cases} 1 \text{, if } j\in S(I(\{i\})) \\ 0\text{, otherwise}
    \end{cases}
\end{equation*}
which deals with our conditions. Finally, we can write down the optimisation problem as  
\begin{align*}
\max_{\boldsymbol{O}} \quad & \boldsymbol{l}^t\boldsymbol{O}+\frac{1}{2}\boldsymbol{O}^tM\boldsymbol{O}\\
\textrm{s.t.} \quad & G\boldsymbol{O} \leq \boldsymbol{h}\\
  &\boldsymbol{O}\geq 0    \\
\end{align*}

\section*{Optimiser in the homogeneous case}

Suppose that the population is homogeneous in foraging ability. Meaning, $N_k=N>0$ for $k\in [\mathcal{N}]$.

\begin{proposition}
    Suppose $N_k=N>0$ $\forall k\in [\mathcal{N}]$. Let $\hat{f}\colon [\mathcal{N}]\times [\mathcal{N}]\to \mathbb{N}$ be defined by 
    \begin{equation*}
        \hat{f}(i,j) =
            |I^{-1}(i)\cap I^{-1}(j)|
    \end{equation*}
    Then all of the coefficients in the objective function $T(\mathbf{O})$ can be expressed as
    \begin{align*}
        & l_i =N \sum_{k=2}^{f(i)}{f(i) \choose k} kN^{-k}, \\
        %& M_{ii} = -2\sum_{k=2}^{f(i)}{f(i) \choose k}kN^{-k}, \\
        & M_{ij} = m_{ij,1}+m_{ij, 2}, \text{where:} \\
        & \hspace{40pt} m_{ij, 1} = \begin{cases}
            -2\sum_{k=2}^{\hat{f} (i, j)}{\hat{f}(i, j)\choose k}kN^{-k}\text{, if }\hat{f}(i,j) \geq 2   \\ 0, \text{ otherwise} 
        \end{cases} , \\ 
        & \hspace{40pt} m_{ij, 2} = \begin{cases}
            -\sum_{k=1}^{\hat{f}(i,j)}{\hat{f}(i,j)\choose k}(f(j)+k-1)N^{-f(i)-k} \text{, if } I^{-1}(j) \subset I^{-1}(i) \\ 
            0\text{, otherwise}
        \end{cases}\\
    %     \begin{cases}
    % -2\sum_{k=2}^{\hat{f} (i, j)}{\hat{f}(i, j)\choose k}kN^{-k} -\sum_{k=1}^{f(i)-f(j)} {f(i)-f(j) \choose k}(f(j)+k-1)N^{-(f(j)+k)}, \text{ if }\hat{f}(i,j) \geq 3 \\ , \text{ if }\hat{f}(i,j) = 2 \\
    % 0 , \text{ else }
    %     \end{cases},\\
        &M_{ji} = M_{ij}
    \end{align*}
    where $i,j\in [\mathcal{N}]$ and $i\geq j$. 
    \begin{proof}
        First we deal with the linear terms given in the vector $\boldsymbol{l}=(l_i)$. In generality, these are given by
        \begin{equation*}
            l_i= \sum_{c\in J(i)} \left[\frac{\sum_{k\in I^{-1}(c)}N_k}{\prod_{k\in I^{-1}(c)}N_k}\right]
        \end{equation*}
        which under the homogeneity assumption (that $N_k=N$ $\forall c\in [\mathcal{N}]$), this simplifies to: 
        \begin{align*}
            l_i= \sum_{c\in J(i)} \left[\frac{f(c)N}{N^{f(c)}}\right] = \sum_{c\in J(i)} f(c){N^{1-f(c)}}. 
        \end{align*}
        The set $J(i)$ is the collection of indices $c\in[\mathcal{N}]$ such that the set $I^{-1}(c)\subseteq I^{-1}(i)$. Since $f(i)$ is the size of the set $I^{-1}(i)$, there are $f(i)\choose k$ many subsets of $I^{-1}(i)$ of size $k$ for $k=2\dots, f(i)$. Since $l_i$ depends only upon the size of the subsets under the homogeneity assumption, we can group terms in the sum by their size. Doing this, we obtain the required expression
        \begin{equation*}
            l_i = \sum_{k=2}^{f(i)}kN^{1-k} = N\sum_{k=2}^{f(i)}kN^{-k}.
        \end{equation*}

        We first compute $m_{ij, 1}$. This has the form:
        \begin{equation*}
            m_{ij, 1} = -\sum_{c\in V(i,j)}\frac{2f(c)}{\prod_{k\in I^{-1}(c)}N_k} = -2\sum_{c\in V(i,j)}f(c)N^{-f(c)}
        \end{equation*}
        The set $V(i,i)$, by definition, is given by 
        \begin{equation*}
            V(i,j) = \left\{c: I^{-1}(c)\subseteq I^{-1}(i)\cap I^{-1}(j) \right\}.
        \end{equation*}
        The size of the set $I^{-1}(i)\cap I^{-1}(j)$ is given by $\hat{f}(i,j)$ by definition. If $\hat{f}(i,j)=0$ or $\hat{f}(i,j)=1$, then $V(i,j)$ should be empty since we do not consider subgroups of size 0 or 1. If $\hat{f}(i,j)\geq 2$, then we can group terms in the sum by their corresponding subgroup size again, as done with the linear coefficients:
        \begin{equation*}
            m_{ij, 1}=  - 2\sum_{k=2}^{\hat{f}(i,j)}{\hat{f}(i,j)\choose k}kN^{-k}.
        \end{equation*}

        Now we focus on $m_{ij,2}$. In the previous subsection, this is denoted by $-\operatorname{Coeff}_2(O_iO_j)$. Since $i\geq j$, we cannot have that $I^{-1}(i)\subset I^{-1}(j)$ by the definition of the index map $I$. If $I^{-1}(j)\not\subset I^{-1}(i)$ then $m_{ij,2}=0$. If $I^{-1}(j)\subset I^{-1}(i)$, then
        \begin{equation*}
            m_{ij,2} = -\sum_{c\in Z(i: j)}\frac{g(j)}{\prod_{k\in I^{-1}(c)}N_k} = -\sum_{c\in Z(i: j)}g(j)N^{-f(c)}.
        \end{equation*}
        The set $Z(i:j)$, by definition, is given by
        \begin{equation*}
            Z(i:j) = \{c:I^{-1}(j) \subset I^{-1}(c) \subseteq I^{-1}(i) \}
        \end{equation*}
        and therefore gives the number of `intermediate' sets between $I^{-1}(j)$ and $I^{-1}(i)$, including $I^{-1}(i)$ but excluding $I^{-1}(j)$. We therefore can break subsets again into their size, by noting that the number of intermediate sets of size will contain $f(j)$ and some additional elements from $f(i)$ not already contained in $f(j)$. The number of sets with $k$ additional elements will be ${f(i)-f(j)\choose k}$ for $k=1, \dots, f(i)-f(j)$. These sets will be of size $f(j)+k$. Therefore, in the case of $I^{-1}(j)\subset I^{-1}(i)$, we can write:
        \begin{equation*}
            m_{ij,2} = -\sum_{k=1}^{f(i)-f(j)}{
            f(i)-f(j)\choose k}(f(j)+k-1)N^{-f(j)-k}.
        \end{equation*}
        Then observing that $\hat{f}(i,j) = f(i) - f(j)$ under the condition $f(i)-f(j)$, we can express
        \begin{equation*}
            m_{ij,2} = -\sum_{k=1}^{\hat{f}(i,j)}{\hat{f}(i,j)\choose k}(f(j)+k-1)N^{-f(j)-k}.
        \end{equation*}
        Then $M_{ij} = m_{ij,1}+m_{ij,2}$ for $i\geq j$, giving all of the lower-diagonal entries. By the symmetry of $M$ we have all coefficients of $T$ simplified using the homogeneity assumption.

    \end{proof}
\end{proposition}

%Define omega as feasible region, and omega_O_^* as the feasible pertubation region. 

\begin{definition}
    The \textit{feasible region} is the set
    \begin{equation*}
        \Omega = \{\boldsymbol{O}\in \mathbb{R}^\mathcal{N} : G\boldsymbol{O}\leq \boldsymbol{h}, \boldsymbol{O} \geq 0\}.
    \end{equation*}
    Meaning, the set of points in which the constraints of the problem (as given in the previous section) are not violated. 
\end{definition}

\begin{definition}
    The \textit{feasible perturbation region about a point} $\boldsymbol{O}\in\mathbb{R}^\mathcal{N}$ is the set \begin{equation*}
        \Omega_\varepsilon(\boldsymbol{O}) = \{\boldsymbol{\varepsilon}\in\mathbb{R}^\mathcal{N}: \boldsymbol{O}+\boldsymbol{\varepsilon}\in \Omega\} 
    \end{equation*}
    Meaning, the set of perturbations from the point $\boldsymbol{O}$ which remain in the feasible set.  
\end{definition}

\begin{lemma}\label{lemma: optim cond}
    A feasible point $O^*\in\Omega$ is a local maximiser of $T$ if and only if there exists a neighbourhood $U\in \Omega_{\varepsilon}({O^*})$ around $\mathbf{0}$ such that every $ \varepsilon \in U$ satisfies
    \begin{equation} \label{eq: optimisation condition}
        T(\varepsilon) \leq -O^*M\varepsilon.
    \end{equation}
    Furthermore, if this property holds for $U= \Omega_{\varepsilon}({O^*})$, then $O^*$ is a global maximiser of $T$. We refer to this inequality as the optimality condition for $O^*$ with perturbation $\varepsilon$.

    \begin{proof}
        The condition for a feasible point $O^*\in\Omega$ to be a local maximiser of $T$ (\textit{the optimality condition}) is that there exists some $\epsilon>0$ such that $T(O) \leq  T(O^*)$ for all $O\in \Omega$ satisfying $\|O-O^*\|\leq \epsilon$.
        
        Any $O\in \Omega$ can be written as $O = O^* + \varepsilon$ for some $\varepsilon\in \Omega_\varepsilon(O^*)$, since both $O$ and $O^*$ are in $\Omega$ and by the definition of $\Omega_\varepsilon(O^*)$. 
        Using this expression for $O$, we can write $\|O-O^*\| = \|\varepsilon\|$. Therefore, the optimality condition can be rewritten as follows. A feasible point $O^*\in\Omega$ is a local maximiser of $T$ if there exists some $\epsilon>0$ such that $T(O^*+\varepsilon) \leq  T(O^*)$ for all $\varepsilon\in \Omega_\varepsilon(O^*)$ satisfying $\|\varepsilon\|\leq \epsilon$. This is equivalent to the point $O^*$ being optimal if there exists some neighbourhood $U\in \Omega_{\varepsilon}({O^*})$ around $\mathbf{0}$ such that every $ \varepsilon \in U$ satisfies
        $T(O^*+\varepsilon) \leq  T(O^*)$. Then observe that 
        \begin{align*}
            T(O^*+\varepsilon) &= \frac{1}{2}(O^{*t}+\varepsilon^t) M (O^{*}+\varepsilon) + l^t( O^{*}+\varepsilon) \\ 
            &=\frac{1}{2}O^{*t}MO^* + \frac{1}{2}O^{*t}M\varepsilon+\frac{1}{2}\varepsilon^tMO^* + \frac{1}{2}\varepsilon^tM\varepsilon + l^tO^{*} + l^t\varepsilon \\
            &=T(O^*) + T(\varepsilon) + \frac{1}{2}O^{*t}M\varepsilon+\frac{1}{2}\varepsilon^tMO^* \\ 
            &= T(O^*) + T(\varepsilon)+O^{*t}M\varepsilon, \text{ since } M \text{ is symmetric, so } O^{*t}M\varepsilon=\varepsilon^tMO^*.
        \end{align*}
        This implies the equivalence
        \begin{align*}
            T(O^* + \varepsilon) \leq T(O^*) & \iff T(O^* + \varepsilon)- T(O^*) \leq 0 \\ 
            & \iff T(\varepsilon)+O^{*t}M\varepsilon \leq 0 \\ & \iff T(\varepsilon) \leq -O^{*t}M\varepsilon. 
        \end{align*}
        Therefore, the optimality condition is equivalent to the statement that there exists some neighbourhood $U\in \Omega_{\varepsilon}({O^*})$ around $\mathbf{0}$ such that every $ \varepsilon \in U$ satisfies: 
        \begin{align*}
            T(\varepsilon) \leq -O^{*t}M\varepsilon. 
        \end{align*}
        Replacing $U$ with $\Omega_\varepsilon(O^*)$ throughout analogously gives the condition for global optimality. 
    \end{proof}
\end{lemma}

\begin{lemma}\label{lemma: first perturb}
Let $O^*=(0, \dots, 0, \frac{N}{2})$ and $\varepsilon=(\varepsilon_1,\dots, \varepsilon_\mathcal{N})\in \Omega_{O^*}$ be such that $\varepsilon_i\geq 0$  $\forall i \in [\mathcal{N}]$. Then $\varepsilon$ satisfies $T(\varepsilon) \leq -O^{*t}M\varepsilon$.
\begin{proof}
    Let $\varepsilon=(\varepsilon_i)\in\Omega_\varepsilon (O^*)$ be any perturbation of the allowed form, where $\varepsilon_i \geq 0$ for $i=1,\dots, \mathcal{N}$. Define $v=(v_i) = O^{*t}M$. Then observe
    \begin{align*}
        v_i &= \sum_{j=1}^\mathcal{N} M_{ji}O_j \\
        &= M_{\mathcal{N}i}\frac{N}{2} \\
        &=\left(-2\sum_{k=2}^{f(i)}{f(i)\choose k}kN^{-k} -\sum_{k=1}^{n-f(i)}{n-f(i) \choose k}(f(i)+k-1)N^{-f(i)-k}\right) \frac{N}{2} \\
        & = -N\sum_{k=2}^{f(i)}{f(i)\choose k}kN^{-k} -\frac{N}{2}\sum_{k=1}^{n-f(i)}{n-f(i) \choose k}(f(i)+k-1)N^{-f(i)-k} \\ & = -l_i - \frac{N}{2}\sum_{k=1}^{n-f(i)}{n-f(i) \choose k}(f(i)+k-1)N^{-f(i)-k} .
    \end{align*}
    Therefore, $v_i\leq -l_i$, since the second term in the right hand side of the final expression is strictly negative. This implies that 
    \begin{equation*}
        \sum_{i=1}^\mathcal{N} l_i\varepsilon_i \leq -\sum_{i=1}^\mathcal{N}{v_i}\varepsilon_i.
    \end{equation*}
    Therefore, since all of the elements of $M$ are non-positive, we also have that
    \begin{equation*}
        \frac{1}{2}\sum_{i=1}^\mathcal{N}\sum_{j=1}^\mathcal{N}\varepsilon_i\varepsilon_jM_{ij} + \sum_{i=1}^\mathcal{N} l_i\varepsilon_i \leq -\sum_{i=1}^\mathcal{N}{v_i}\varepsilon_i.
    \end{equation*}
    Which is equivalent to
    \begin{equation*}
        T(\varepsilon) \leq -v^t\varepsilon=-O^{*t}M\varepsilon
    \end{equation*}
    as required.
    
\end{proof}
\end{lemma}

\begin{lemma}\label{lemma: second perturb}
Let $O^*=(0, \dots, 0, \frac{N}{2})$ and $\varepsilon=(0,\dots, 0, -\varepsilon_\mathcal{N})$ be such that $\varepsilon_\mathcal{N}>0$. Then $\varepsilon$ satisfies $T(\varepsilon) \leq -O^{*t}M\varepsilon$.  
\begin{proof}
    Let $\varepsilon=(\varepsilon_i)\in\Omega_\varepsilon (O^*)$ be any perturbation of the allowed form, where $\varepsilon_i = 0$ for $i=1,\dots, \mathcal{N}-1$ and $\varepsilon_\mathcal{N} < 0$. Observe the following equivalence: 
    \begin{align*}
        T(\varepsilon) \leq -O^{*t}M\varepsilon & \iff \frac{1}{2}\varepsilon^{t}M\varepsilon+l^t\varepsilon \leq -O^{*t}M\varepsilon \\ 
        &\iff \sum_{i=1}^\mathcal{N}\sum_{j=1}^\mathcal{N} \varepsilon_iM_{ij}\varepsilon_j + \sum_{i=1}^\mathcal{N}\varepsilon_il_i \leq -\sum_{i=1}^\mathcal{N}\sum_{j=1}^\mathcal{N}O^*_iM_{ij}\varepsilon_j 
        \\& \iff \varepsilon_\mathcal{N}^2M_{\mathcal{N}\mathcal{N}}+ \varepsilon_\mathcal{N}l_\mathcal{N} \leq -O^*_\mathcal{N} M_{\mathcal{N}\mathcal{N}}\varepsilon_\mathcal{N} 
        \\& \iff \varepsilon_\mathcal{N}M_{\mathcal{N}\mathcal{N}}+ l_\mathcal{N} \geq -O^*_\mathcal{N} M_{\mathcal{N}\mathcal{N}} 
        \\&
        \iff \left(\varepsilon_\mathcal{N} +O^*\right)M_{\mathcal{N}\mathcal{N}} + l_\mathcal{N} \geq 0 
        \\&
        \iff -2\left(\varepsilon_\mathcal{N}+\frac{N}{2}\right) \left(\sum_{k=2}^n {n \choose k}kN^{-k} \right)+ N\left(\sum_{k=2}^n{n \choose k}kN^{-k}\right) \geq 0 \\ 
        &\iff -2\varepsilon_\mathcal{N} -  N +  N\geq 0 \\
        & \iff \varepsilon_\mathcal{N} \leq 0
    \end{align*}
    which is true by assumption.
\end{proof}
\end{lemma}

\begin{remark}
    Any $\varepsilon \in \Omega_{O^*}$ can be written in the form of $\varepsilon^{(1)}$, as defined in Lemma \ref{lemma: first perturb}, or in the form of $\varepsilon^{(1)} + \varepsilon^{(2)}$, where $\varepsilon^{(2)}$ is as defined in Lemma \ref{lemma: second perturb}. These two cases correspond to the distinction between feasible perturbations in the positive and negative directions with respect to the final coordinate $O_\mathcal{N}$, respectively. Other coordinates must be perturbed from $O^*$ in the positive direction, since $O^*$ lies on the boundary of $\Omega$ (as $O_i = 0$ for $i=1,\dots, \mathcal{N}-1$). 
\end{remark}

\begin{theorem}\label{thm: optim}
    Let $O^*=(0, \dots, 0, \frac{N}{2})$. Then $O^*$ is the global maximiser of $T$. 
\begin{proof}
    We construct a neighbourhood $U$ about $\boldsymbol{0}$ in $\Omega_\varepsilon(O^*)$ such that all $\varepsilon\in U$ satisfy
    \begin{equation}\label{eq: optimality}
        T(\varepsilon) \leq -O^{*t}M\varepsilon
    \end{equation}
    which would imply that $O^*$ satisfies the optimality condition of Lemma \ref{lemma: optim cond} in the local sense. We then show that this constructed set must be equal to $\Omega_\varepsilon(O^*)$ itself, so that $O^*$ also satisfies the optimality condition of Lemma \ref{lemma: optim cond} in the global sense. To construct this set, we consider which perturbations in $\Omega_\varepsilon(O^*)$ satisfy inequality \eqref{eq: optimality}. As noted in the previous remark, each feasible perturbation can be written in the form of $\varepsilon^{(1)}$, or in the form of $\varepsilon^{(1)} + \varepsilon^{(2)}$. We consider these two cases separately.

    In the first case, all perturbations are of the form $\varepsilon = (\varepsilon_i)$ and $\varepsilon_i \geq 0$. By Lemma \ref{lemma: first perturb}, the condition \eqref{eq: optimality} holds for any such $\varepsilon\in\Omega_\varepsilon(O^*)$.

    In the second case, perturbations are of the form $\varepsilon = \varepsilon^{(1)} + \varepsilon^{(2)}$. Denote $\varepsilon^{(1)} = (\varepsilon^{(1)}_i)$ and $\varepsilon^{(2)} = (\varepsilon^{(2)}_i)$. Perturbations of this form will have $\varepsilon^{(1)}_i \geq 0$ for $i=1,\dots, \mathcal{N}$, $\varepsilon^{(2)}_j\geq 0$ for $i=1,\dots, \mathcal{N}-1$ and $\varepsilon^{(2)}_\mathcal{N}< 0$. Without loss of generality, assume that $\varepsilon^{(1)}_\mathcal{N} = 0$. Now observe that:
    \begin{align*}
        T(\varepsilon) &=   T\left(\varepsilon^{(1)}+ \varepsilon^{(2)}\right) \\
        & = \frac{1}{2}\left(\varepsilon^{(1)} + \varepsilon^{(2)}\right)^t M \left(\varepsilon^{(1)} + \varepsilon^{(2)}\right) + l^t\left(\varepsilon^{(1)} + \varepsilon^{(2)}\right) \\
        &=T\left(\varepsilon^{(1)}\right) + T\left(\varepsilon^{(2)}\right) + \varepsilon^{(1)t}M\varepsilon^{(2)} \\ 
        & \leq T(\varepsilon^{(1)})-O^{*t}M\varepsilon^{(2)} + \varepsilon^{(1)t}M\varepsilon^{(2)}\text{, by Lemma \ref{lemma: second perturb}} .
    \end{align*}
    Furthermore, by Lemma \ref{lemma: first perturb}, we have that
    \begin{equation*}
        T\left(\varepsilon^{(1)}\right) \leq -O^*M\varepsilon^{(1)},
    \end{equation*}
    which implies that 
    \begin{equation}\label{eq: X}
        X\vcentcolon= -T\left(\varepsilon^{(1)}\right)-O^*M\varepsilon^{(1)}
    \end{equation}
    is non-negative. Our current inequality for $T(\varepsilon)$ can be expressed as
    \begin{align*}
        T(\varepsilon) \leq -O^{*t}M\varepsilon^{(1)}-X-O^{*t}M\varepsilon^{(2)} + \varepsilon^{(1)t}M\varepsilon^{(2)}, 
    \end{align*}
    which implies that
    \begin{equation*}
        T(\varepsilon) \leq  -O^{*t}M\varepsilon +\left(-X +\varepsilon^{(1)t}M\varepsilon^{(2)}\right).
    \end{equation*}
    Therefore, $\varepsilon$ will satisfy the optimality condition \eqref{eq: optimality} if 
    \begin{equation*}
        -X +\varepsilon^{(1)t}M\varepsilon^{(2)} \leq 0 \iff X \geq \varepsilon^{(1)t}M\varepsilon^{(2)}.
    \end{equation*}
    Expanding the $\varepsilon^{(1)t}M\varepsilon^{(2)}$ term gives:
    \begin{equation*}
        \varepsilon^{(1)t}M\varepsilon^{(2)} = \sum_{i=1}^\mathcal{N}\sum_{j=1}^\mathcal{N}\varepsilon_j^{(1)}M_{ij}\varepsilon^{(2)}_i = \sum_{j=1}^\mathcal{N} \varepsilon_j^{(1)}M_{\mathcal{N}j}\varepsilon_\mathcal{N} = \varepsilon_\mathcal{N}\sum_{j=1}^\mathcal{N} \varepsilon_j^{(1)}M_{\mathcal{N}j}.
    \end{equation*}
    Therefore, $\varepsilon$ will satisfy the optimality condition \eqref{eq: optimality} if 
    \begin{equation*}
        X \geq  \varepsilon_\mathcal{N}\sum_{j=1}^\mathcal{N} \varepsilon_j^{(1)}M_{\mathcal{N}j} \implies \varepsilon_\mathcal{N} \geq \frac{X}{\sum_{j=1}^\mathcal{N} \varepsilon_j^{(1)}M_{\mathcal{N}j}},
    \end{equation*}
    using the fact that all entries of $M$ are negative and all elements of $\varepsilon^{(1)}$ are positive, which implies the above sum is negative. We use this property to construct the set $U$.
    
    Consider the map $P\colon \Omega_\varepsilon(O^*) \to \mathbb{R}^\mathcal{N}$ which maps the final coordinate to zero and fixes the remaining coordinates. Define $U'= P(\Omega_\varepsilon(O^*))$ and
    \begin{equation*}
        K\vcentcolon= \min_{\varepsilon\in U'}\left[\frac{X}{\sum_{j=1}^\mathcal{N} \varepsilon_j^{(1)}M_{\mathcal{N}j}} \right],
    \end{equation*}
    which must exist, be finite and be non-zero since $N\neq 0$. Using the definition of $X$ we can write
    \begin{align*}
        K &= \min_{\varepsilon\in U'}\left[\frac{-T\left(\varepsilon^{(1)}\right)-O^*M\varepsilon^{(1)}}{\sum_{j=1}^\mathcal{N} \varepsilon_j^{(1)}M_{\mathcal{N}j}} \right] \\
        &  = \min_{\varepsilon\in U'}\left[\frac{-T\left(\varepsilon^{(1)}\right)-\frac{N}{2}\sum_{j=1}^\mathcal{N}\varepsilon_j^{(1)}M_{\mathcal{N}j}}{\sum_{j=1}^\mathcal{N} \varepsilon_j^{(1)}M_{\mathcal{N}j}} \right] \\ 
        & = \min_{\varepsilon\in U'}\left[\frac{-T\left(\varepsilon^{(1)}\right)}{\sum_{j=1}^\mathcal{N} \varepsilon_j^{(1)}M_{\mathcal{N}j}} \right] - \frac{N}{2}
    \end{align*}
    Now observe that the quantity in the minimisation operator can take the value 0, exactly when $T(\varepsilon) = 0$. We can construct a value of $\varepsilon\in U'\setminus \{0\}$ which solves $T(\varepsilon)$ as follows. Split $[n]$ into two \textit{disjoint} subsets, $S_1$ and $S_2$, consisting of $\left\lceil \frac{n}{2} \right\rceil$ and $\left\lfloor \frac{n}{2} \right\rfloor$ many individuals, respectively. Set $\varepsilon_j = N$ for $j = I(S_1)$ and $j = I(S_2)$, and all other coordinate values $\varepsilon_i=0$. Therefore, for all $c\in \left[\mathcal{N}\right]$, we have that either $P(c)=0$ (for the zero overlaps) or $A(c)=0$ (for the two non-zero overlaps of individuals, where the amount of unique knowledge is zero), which implies that the net information transfer, $T(\varepsilon)$, is $0$. Therefore, we can say 
    \begin{equation*}
        K \leq -\frac{N}{2}
    \end{equation*}
    Then, we can define $U_\mathcal{N}$ as
    \begin{equation*}
        U_\mathcal{N} = \{0\}\times \dots \times \{0\}\times \left[-\frac{N}{2}, \frac{N}{2}\right]
    \end{equation*}
    and then define $U$ as
    \begin{equation*}
        U =\left( U' + U_\mathcal{N}\right) \cap \Omega_{\varepsilon}(O^*)
    \end{equation*}
    where the $+$ denotes the Minkowski sum of sets (i.e the element-wise sum). By construction, $U$ is a neighbourhood in $\Omega_{\varepsilon}(O^*)$ about $\boldsymbol{0}$ where the condition 
    \begin{equation}
        T(\varepsilon) \leq -O^{*t}M\varepsilon
    \end{equation}
    holds for all $\varepsilon\in U$. Therefore, by Lemma \ref{lemma: optim cond}, $O^*$ is a local optimiser of $T$. For global optimality, it remains to show that $U = \Omega_{\varepsilon}(O^*)$. We do this by showing that 
    \begin{equation*}
        \Omega_{\varepsilon}(O^*) \subseteq U'+U_\mathcal{N}
    \end{equation*}
    which would imply that $U=\Omega_{\varepsilon}(O^*)$. Let $\varepsilon$ be any feasible perturbation. We show that this implies $\varepsilon\in U'+U_\mathcal{N}$. Write $\varepsilon = \varepsilon^{(1)}+\varepsilon^{(2)}$, where $\varepsilon^{(1)}$ contains the first $\mathcal{N}-1$ entries of $\varepsilon$ but has $\varepsilon^{(1)}_\mathcal{N}=0$, and $\varepsilon^{(2)}$ has $\varepsilon^{(2)}_i=0$ for all $i\in[\mathcal{N}]$ except for $\mathcal{N}$. We therefore note that, by definition, $\varepsilon^{(1)}=P(\varepsilon)$, so that $\varepsilon^{(1)}$ is in the image $ P(\Omega_\varepsilon (O^*))$. We show that $\varepsilon^{(2)}$ must be in $U_\mathcal{N}$. This is a straightforward argument. If $\varepsilon_\mathcal{N} < -\frac{N}{2}$, then $O^*+\varepsilon$ does not satisfy the constraint $\boldsymbol{O}\geq 0$. So, no $\varepsilon\in\Omega_\varepsilon (O^*)$ can have this property. So, on the other hand, if $\varepsilon_{\mathcal{N}} > \frac{N}{2}$, then the condition $G(O^*+\varepsilon)\leq h$ can not be satisfied since
    \begin{align*}
    (O^*+\varepsilon)_\mathcal{N} = \frac{N}{2}+\varepsilon_\mathcal{N} > N,
    \end{align*}
    such that no $\varepsilon\in\Omega_\varepsilon (O^*)$ can have this property either. Hence, $\varepsilon^{(2)}\in U_\mathcal{N}$. Therefore, $\varepsilon\in U'+U_\mathcal{N}$. This implies that
    \begin{equation*}
        U = \left( U' + U_\mathcal{N}\right) \cap \Omega_{\varepsilon}(O^*) = \Omega_{\varepsilon}(O^*)
    \end{equation*}
    which therefore proves that $O^*$ is the global optimiser of $T$ over $\Omega$. 
\end{proof}
\end{theorem}

\begin{remark}
    At the global optimiser $O^*=(0, \dots, 0, \frac{N}{2})$, the objective function takes value
    \begin{equation*}
        T(O^*) = \frac{N^2}{2}\sum_{k=2}^n {{n}\choose{k}}kN^{-k} = \frac{n(n-1)}{4} + \mathcal{O}(N^{-1})
    \end{equation*}
    and therefore, when the `number of points' is very large, the optimal value of the information transfer function is approximately quadratic in the number of individuals in the group. 
\end{remark}

\begin{definition}
    The \textit{total relative overlap of the group} of $n$ individuals, $w_n$, is defined as the ratio of the number of cells known by \textit{all} individuals to the number of cells known by at least one individual. In particular:
    \begin{equation*}
        w_n = \frac{O_\mathcal{N}}{\sum_{i=1}^nN_i-\sum_{c=1}^\mathcal{N}g(c)O_c}
    \end{equation*}
    or, equivalently, the ratio of the area shared by all individuals to the total occupied area. 

    This definition is general to both the homogenous and non-homogenous case. But the following result holds only in the homogenous one.
\end{definition}

\begin{corollary}
    At the optimal rate of information transfer, the total relative overlap of the group of $n$ individuals is given by:
    \begin{equation*}
        w_n^* = \frac{1}{n+1}
    \end{equation*}
    \begin{proof}
        By Theorem \ref{thm: optim}, the optimiser of $T$ is $O^*=(0, \dots, 0, \frac{N}{2})$. The corresponding value of $w_n$ is then given by
        \begin{align*}
            w_n^* &= \frac{O_\mathcal{N}}{\sum_{i=1}^nN_i-\sum_{c=1}^\mathcal{N}g(c)O_c} \\&
             = \frac{N/2}{nN-g(\mathcal{N})N/2}\\
             &=\frac{1}{2n-(n-1)}\text{, cancelling the }N/2 \text{ term} \\
             &= \frac{1}{n+1}.
        \end{align*}
    \end{proof}
\end{corollary}

\section{Results}
\subsection{Correlation between variation in fruit abundance and the filtration complementarity index}

\begin{figure}[H]
    \centering
    %left, bottom, right, top
    \includegraphics[trim={0 80 0 110}, clip, width=0.5\textwidth] {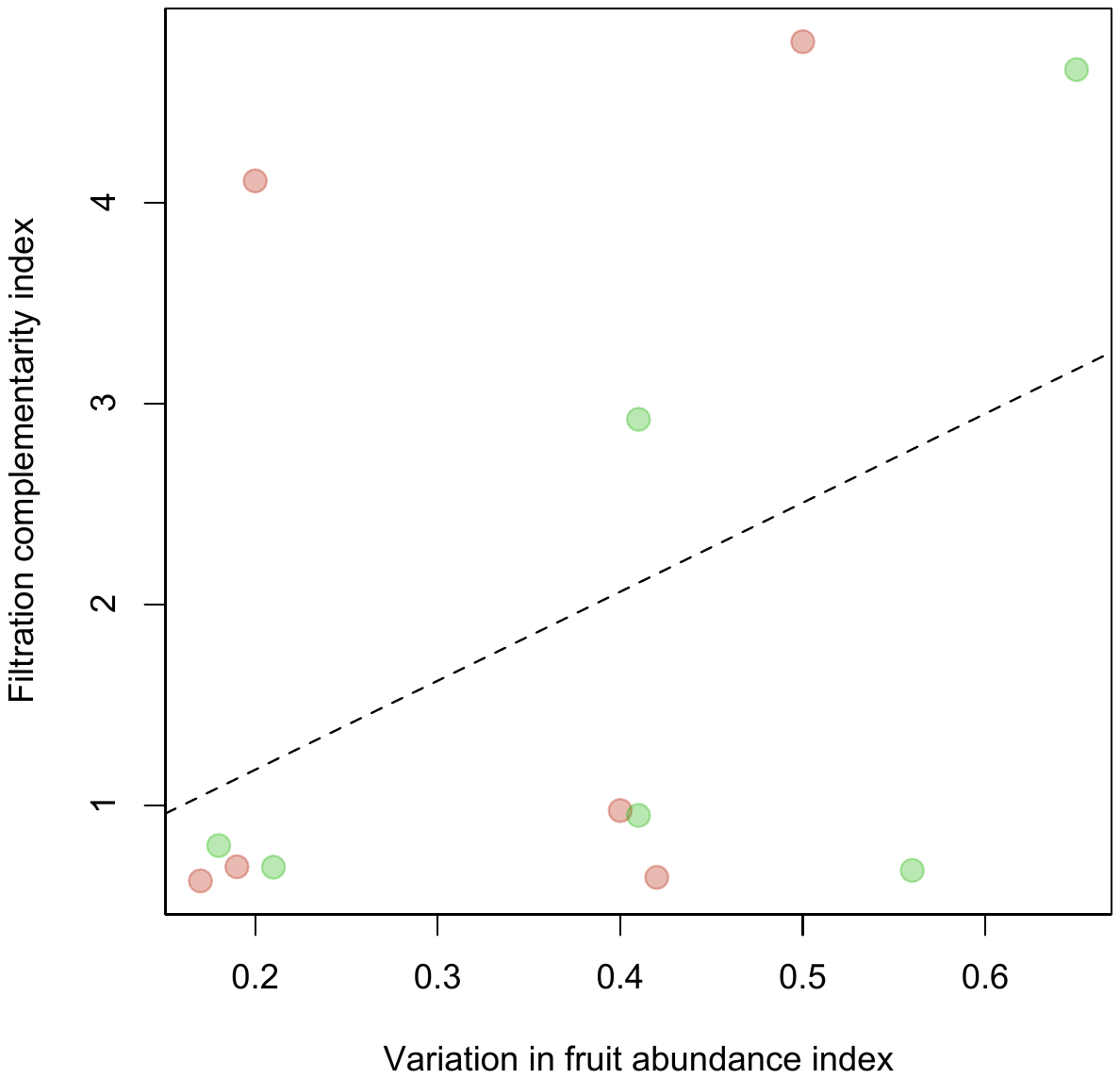}
    \caption{Relationship between the confidence intervals in the biweekly index of fruit abundance and the filtration complementarity index for different seasons. The values of the filtration complementarity index do not vary significantly with season, although they show a tendency in the predicted direction (ANOVA $F_{1}$=2.2, $p=$0.17). Light red: dry seasons; light green: wet seasons. The dotted line corresponds to the non-significant  linear fit to all points, regardless of season ($\rho$ = 0.42 and $P$=0.16).}
    \label{fig:correlation}
\end{figure}

\subsection{Simplicial centrality and other simplex characteristics}

\begin{figure}[H]
    \centering
    \includegraphics[width=1\textwidth]{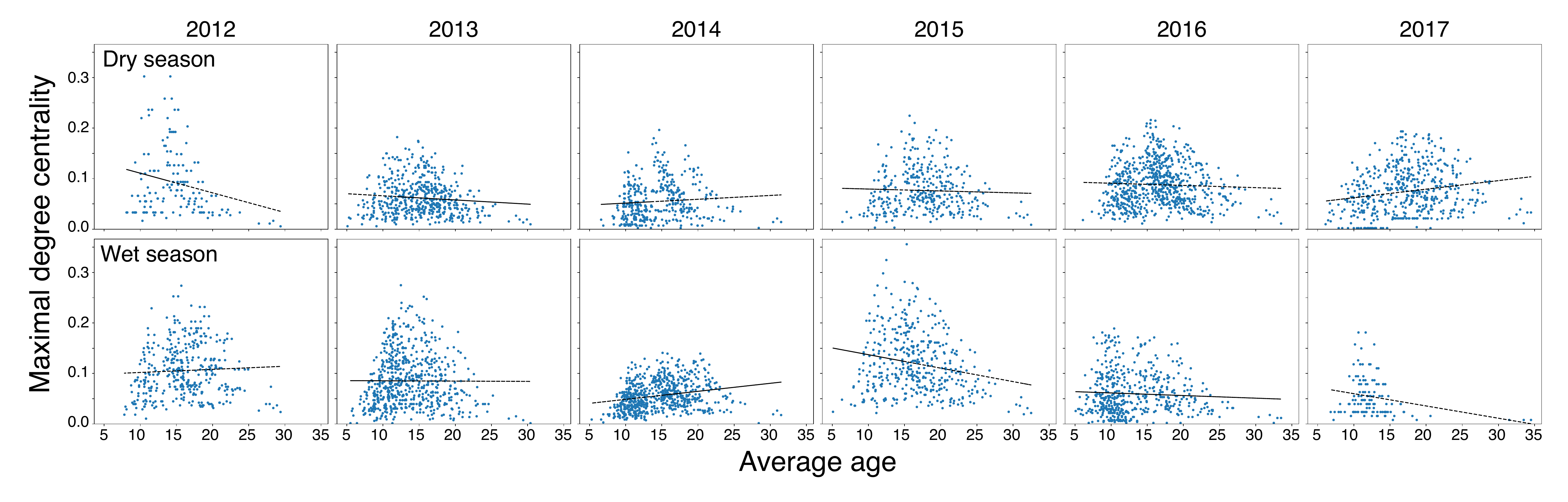}
    \caption{Maximal simplicial degree centrality values for simplices composed of adult individuals of different age. Each dot corresponds to a simplex formed by the intersection of individual core ranges assuming $\alpha \leq 4$. Top row: dry seasons; bottom row: wet seasons.}
    \label{fig:centrality_and_age}
\end{figure}

\begin{figure}[H]
    \centering
    \includegraphics[width=1\textwidth]{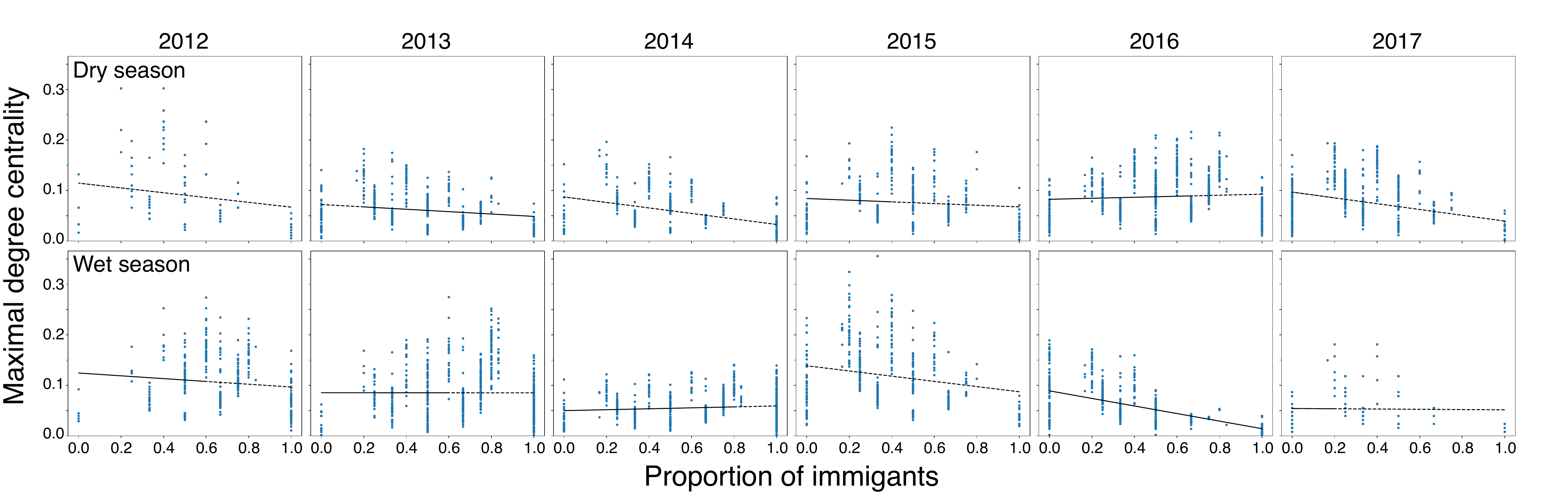}
    \caption{Maximal simplicial degree centrality values for simplices with different proportions of males. Each dot corresponds to a simplex formed by the intersection of individual core ranges assuming $\alpha \leq 4$. Top row: dry seasons; bottom row: wet seasons.}
    \label{fig:centrality_and_sex}
\end{figure}

\begin{figure}[H]
    \centering
    \includegraphics[width=1\textwidth]{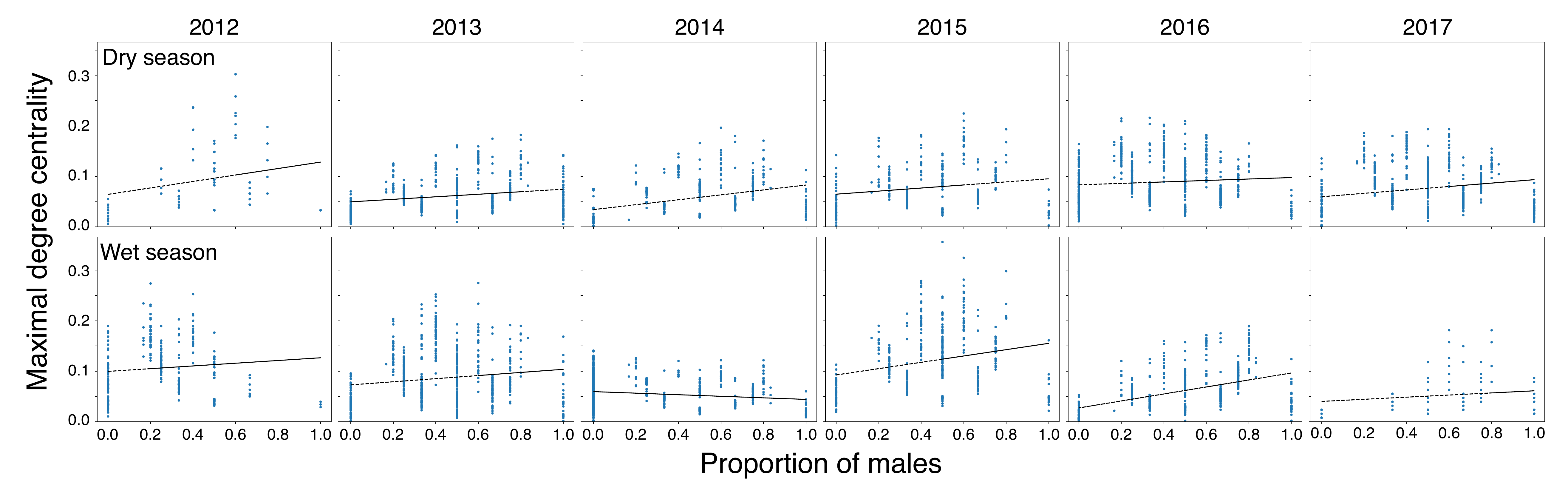}
    \caption{Maximal simplicial degree centrality values for simplices composed of different proportions of recently immigrated females. Each dot corresponds to a simplex formed by the intersection of individual core ranges assuming $\alpha \leq 4$. Top row: dry seasons; bottom row: wet seasons.}
    \label{fig:centrality_and_immigration}
\end{figure}

\subsection{Simplicial centrality and maximum simplex size}

\begin{figure}[H]
    \centering
    \includegraphics[width=1\textwidth]{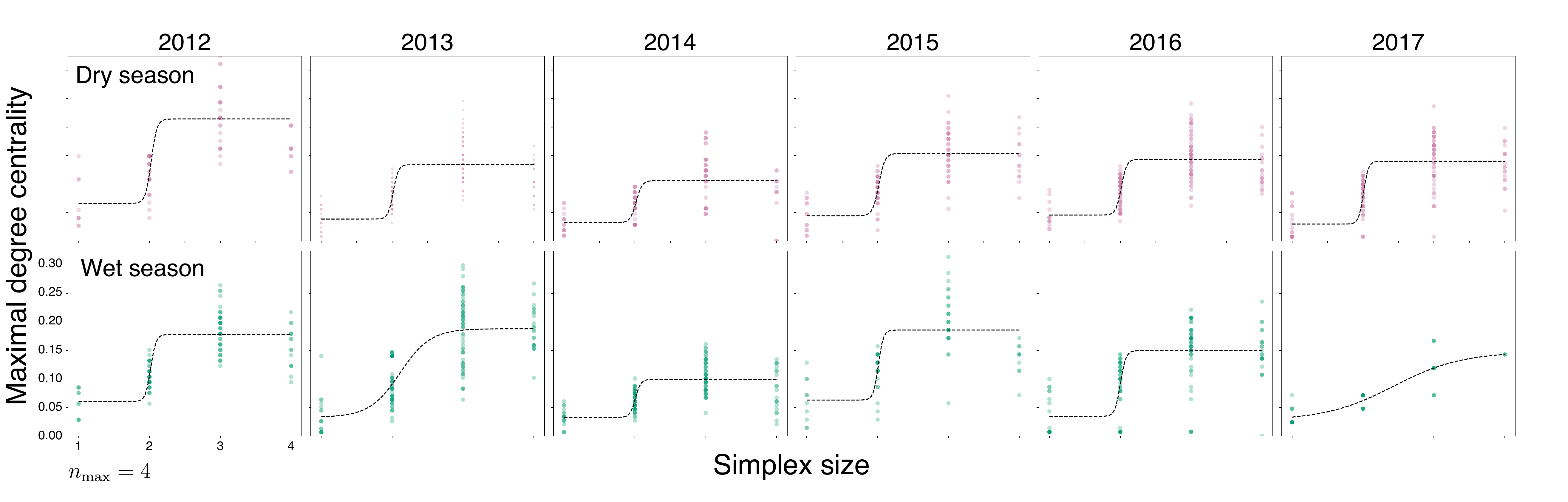}
    \caption{Maximal simplicial degree centrality values for simplices of different size with maximum simplex size parameter $n_\text{max}=4$. Each dot corresponds to a simplex formed by the intersection of individual core ranges assuming $\alpha \leq 4$. Dashed black line corresponds to a continuous sigmoid function fitted to the centrality values using the non-linear least squares method from the \textit{Scipy} package in Python \cite{SciPy2020}. Top row: dry seasons; bottom row: wet seasons.}
    \label{fig:centrality_and_size_n4}
\end{figure}

\begin{figure}[H]
    \centering
    \includegraphics[width=1\textwidth]{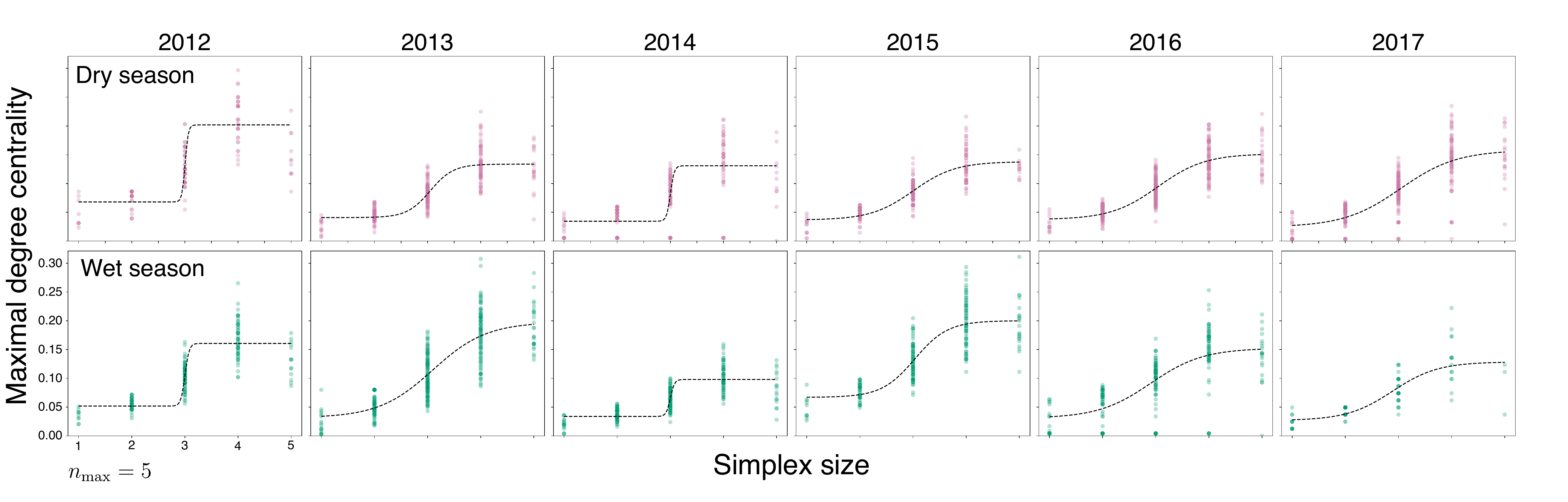}
    \caption{Maximal simplicial degree centrality values for simplices of different size with maximum simplex size parameter $n_\text{max}=5$. Each dot corresponds to a simplex formed by the intersection of individual core ranges assuming $\alpha \leq 4$. Dashed black line corresponds to a continuous sigmoid function fitted to the centrality values using the non-linear least squares method from the \textit{Scipy} package in Python \cite{SciPy2020}. Top row: dry seasons; bottom row: wet seasons.}
    \label{fig:centrality_and_size_n5}
\end{figure}

\begin{figure}[H]
    \centering
    \includegraphics[width=1\textwidth]{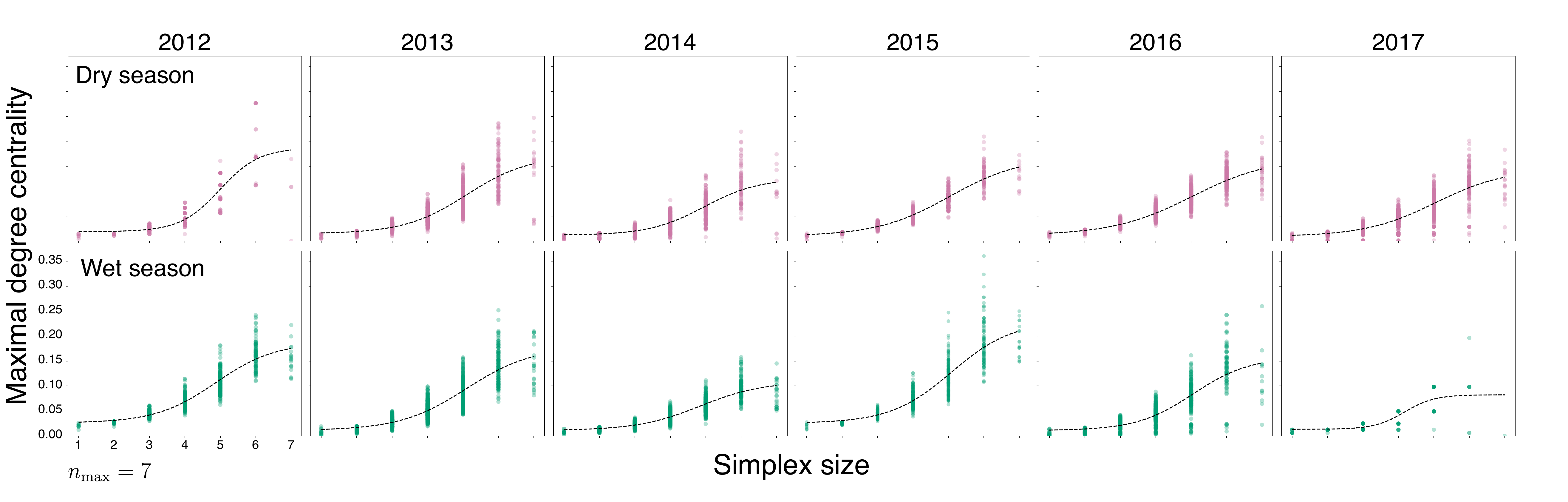}
    \caption{Maximal simplicial degree centrality values for simplices of different size with maximum simplex size parameter $n_\text{max}=7$. Each dot corresponds to a simplex formed by the intersection of individual core ranges assuming $\alpha \leq 4$. Dashed black line corresponds to a continuous sigmoid function fitted to the centrality values using the non-linear least squares method from the \textit{Scipy} package in Python \cite{SciPy2020}. Top row: dry seasons; bottom row: wet seasons.}
    \label{fig:centrality_and_size_n7}
\end{figure}

\begin{figure}[H]
    \centering
    \includegraphics[width=1\textwidth]{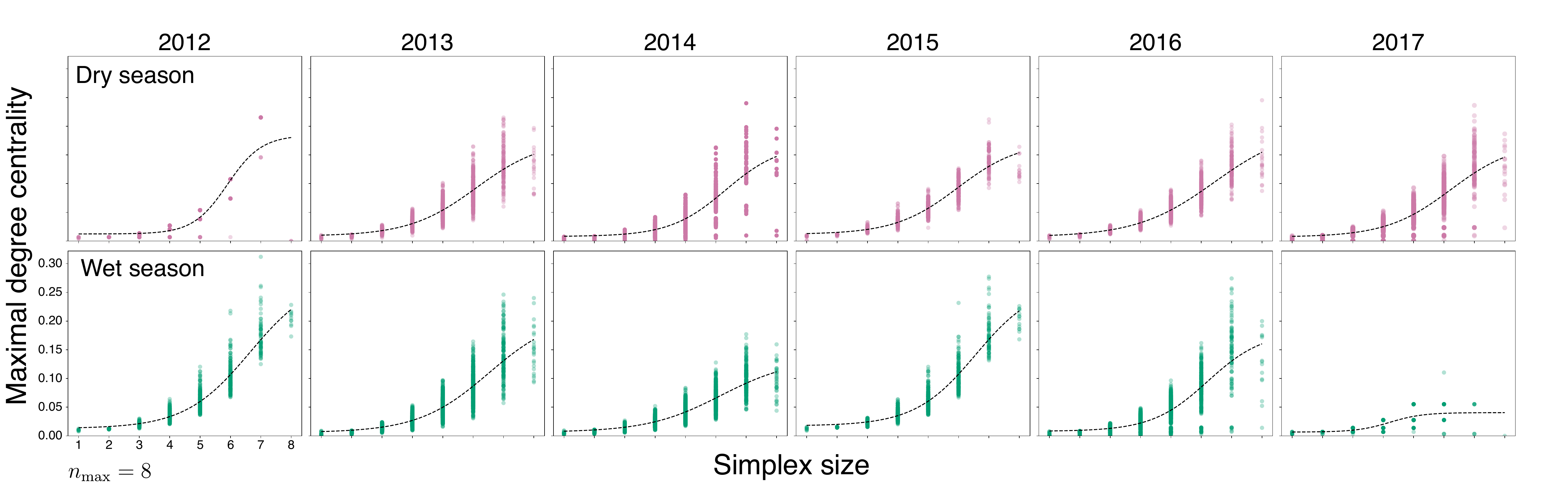}
    \caption{Maximal simplicial degree centrality values for simplices of different size with maximum simplex size parameter $n_\text{max}=8$. Each dot corresponds to a simplex formed by the intersection of individual core ranges assuming $\alpha \leq 4$. Dashed black line corresponds to a continuous sigmoid function fitted to the centrality values using the non-linear least squares method from the \textit{Scipy} package in Python \cite{SciPy2020}. Top row: dry seasons; bottom row: wet seasons.}
    \label{fig:centrality_and_size_n8}
\end{figure}

\begin{figure}[H]
    \centering
    \includegraphics[width=1\textwidth]{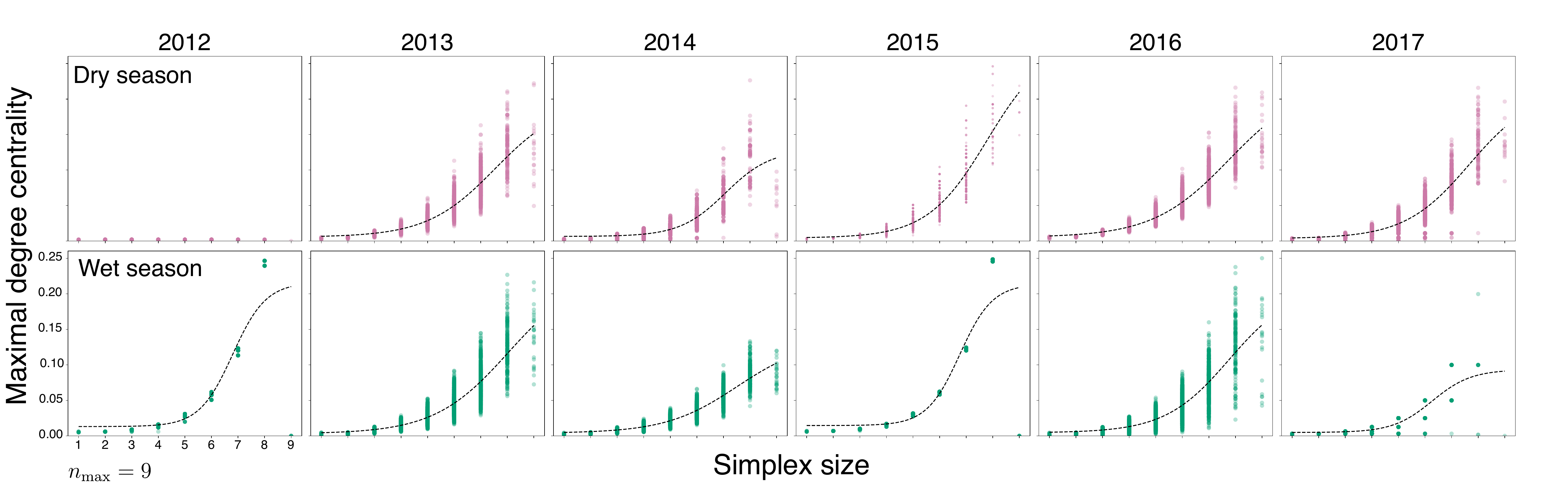}
    \caption{Maximal simplicial degree centrality values for simplices of different size with maximum simplex size parameter $n_\text{max}=9$. Each dot corresponds to a simplex formed by the intersection of individual core ranges assuming $\alpha \leq 4$. Dashed black line corresponds to a continuous sigmoid function fitted to the centrality values using the non-linear least squares method from the \textit{Scipy} package in Python \cite{SciPy2020}. Top row: dry seasons; bottom row: wet seasons. No sigmoid was fitted for the dry season of 2012 due to low centrality values (which arise due to the trivial structure).}
    \label{fig:centrality_and_size_n9}
\end{figure}

\end{document}